\documentclass[11pt,a4paper]{article}		
\pdfoutput=1

\usepackage[utf8]{inputenc}

\usepackage{amsmath}							
\usepackage{amssymb}
\usepackage{graphicx}
\usepackage{color}
\usepackage{authblk}
\usepackage{subfigure}

\newcommand{\rf}[1]{(\ref{#1})}
\newcommand{\beq}{\begin{equation}}
\newcommand{\beql}[1]{\beq\label{#1}}
\newcommand{\eeq}{\end{equation}}
\newcommand{\bea}{\begin{eqnarray}}
\newcommand{\eea}{\end{eqnarray}}

\newcommand{\ra}{\rangle}
\newcommand{\la}{\langle}

\newcommand{\cD}{{\cal D}}

\newcommand{\cM}{{\cal M}}

\newcommand{\cT}{{\cal T}}

\newcommand{\cN}{{\cal N}}

\newcommand{\cO}{{\cal O}}

\newcommand{\cZ}{{\cal Z}}

\definecolor{grn}{rgb}{0,0.85,0}									

\newcommand{\const}{\mathrm{const}}

\newcommand{\dd}{\mathrm{d}}

\begin{document}


\title{Four-dimensional CDT with toroidal topology}

\author[a,b]{J.Ambj\o rn}
\author[c]{J. Gizbert-Studnicki}
\author[a,c]{A. G\"orlich}
\author[a]{K. Grosvenor}
\author[c]{J. Jurkiewicz}

\affil[a]{The Niels Bohr Institute, Copenhagen University\\ Blegdamsvej 17, DK-2100 Copenhagen, Denmark.} 
\affil[b]{IMAPP, Radboud University\\ Nijmegen, PO Box 9010, The Netherlands.}
\affil[c]{The M. Smoluchowski Institute of Physics, Jagiellonian University\\ \L ojasiewicza 11, Krak\'ow, PL 30-348, Poland.}



\maketitle

\begin{abstract}
3+1 dimensional  Causal Dynamical Triangulations (CDT) 
describe a quantum theory of fluctuating geometries without the 
introduction of a background geometry. If the topology of space is constrained
to be that of a three-dimensional torus we show that the system will 
fluctuate around a dynamically formed  background geometry
which can be understood from a simple minisuperspace action which 
contains both a classical part and a quantum part. We 
determine this action by integrating out degrees of freedom
in the full model, as well as by transfer matrix methods.

\end{abstract}

\section{Introduction}

Perturbative methods combined with ``traditional'' quantum field theory (QFT) 
techniques provide a powerful tool in describing three out of the four fundamental 
interactions, gravity being the inglorious exception. 
This is because QFT based on Einstein's general relativity is perturbatively 
nonrenormalizable \cite{tHooft1974} and as such perturbative calculations around any fixed background geometry can at most be treated as an effective theory valid up to some  energy scale, much lower than the Planck energy, at which one expects quantum effects to play an important role. However, there is still hope that one can use the QFT approach in the quest for quantizing gravity, but one has to go beyond the simple perturbative framework. 
This idea is known as the {\it asymptotic safety} conjecture 
\cite{weinberg}.
It  assumes that the renormalization group flow of   the bare gravitational couplings leads to a non-Gaussian UV fixed point at which the  QFT in question is finite and predictive, but where the couplings do not need to be small (thus 
invalidating the naive use of perturbation theory). 
The existence of such a non-trivial UV fixed point is supported by the $\epsilon$-expansion
near two dimensions \cite{kawai} and by the use of so-called
functional renormalization group calculations \cite{FRG}. 

The actual calculations using the functional renormalization group  make approximations 
which can be difficult to control, and it is important to verify the results by independent 
methods. Lattice QFT is a non-perturbative tool which has successfully addressed 
questions in QCD which are beyond perturbation theory. QCD is an ordinary field 
theory in flat spacetime and the hyper-cubic lattice used represents a simple 
discretization of this flat spacetime. However, if the field theory is gravity, spacetime 
itself becomes dynamical. The formalism of dynamical triangulations (DT) provides 
a way to sum over such fluctuating geometries in a path integral approach \cite{DT,higherDT}. It has provided a very successful regularization of  two-dimensional Euclidean quantum gravity coupled to matter fields  theories which can be solved analytically 
both on the lattice and using standard continuum QFT (for reviews see \cite{dgz,book}). 
Taking the scaling limit of 
the lattice theory one recovers the continuum theory. 
Thus, even if it is a lattice theory, it is clearly able to  provide a 
lattice regularization of diffeomorphism  invariant theories. Lattice theories using various 
classes of dynamical triangulations are thus good candidates for lattice theories
which can be used in the path integral formulation of quantum gravity, and if a 
non-trivial UV fixed point exists for quantum gravity one should be able to 
identify it using lattice methods. The starting point would thus be the formal path integral  
\beql{Zcont}
\cZ= \int \cD_{\cM}{[g]} e^{iS_{HE} [g]}\ ,
\eeq
where one integrates over geometries, i.e. over all physically distinct metric tensors $g$, and $S_{HE}$ is some  classical action, e.g. the Hilbert-Einstein action. One then computes expectation values or correlators of  physical observables $\cO_1[g], ... , \cO_n[g]$ as
\beql{Correlcont}
\langle \cO_1[g], ... ,\cO_n[g] \rangle= \int \cD_{\cM}{[g]} e^{iS_{HE} [g]}  \cO_1[g]... \cO_n[g] \ . 
\eeq
In gravity it is a difficult question to find suitable observables, but this problem is 
not linked to the lattice regularization, but is present already in the continuum 
formulation.

This article deals with a particular choice of dynamical triangulations which 
has been denoted Causal Dynamical Triangulations (CDT) \cite{cdt}, and which has shown
an interesting phase diagram, in terms of the bare coupling constants which enter 
into the action. The possibility of having second order phase transitions opens
up the possibility to take a continuum limit and in this way make contact to 
other approaches studying asymptotic safety. We refer to \cite{physrep} for a 
review of the technical implementation of CDT. For completeness
we here provide a short review.

\begin{figure}
	\centering
	\includegraphics[width=0.45\linewidth]{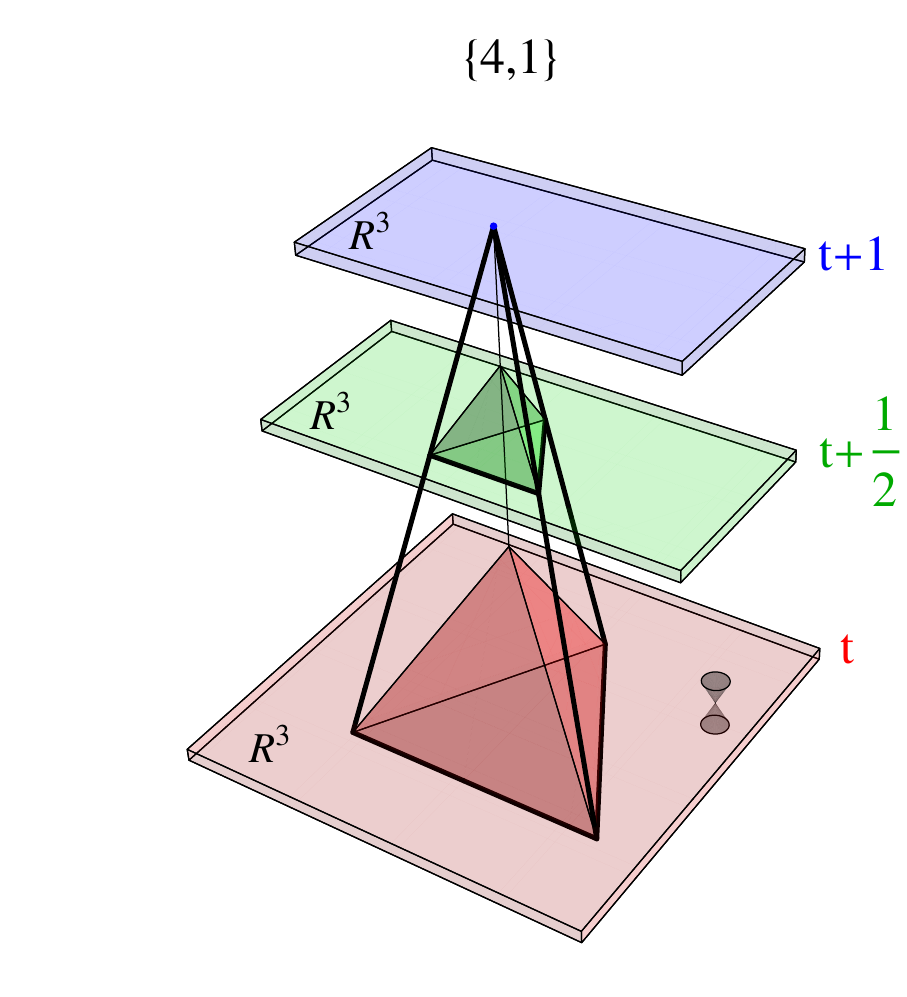}
	\includegraphics[width=0.45\linewidth]{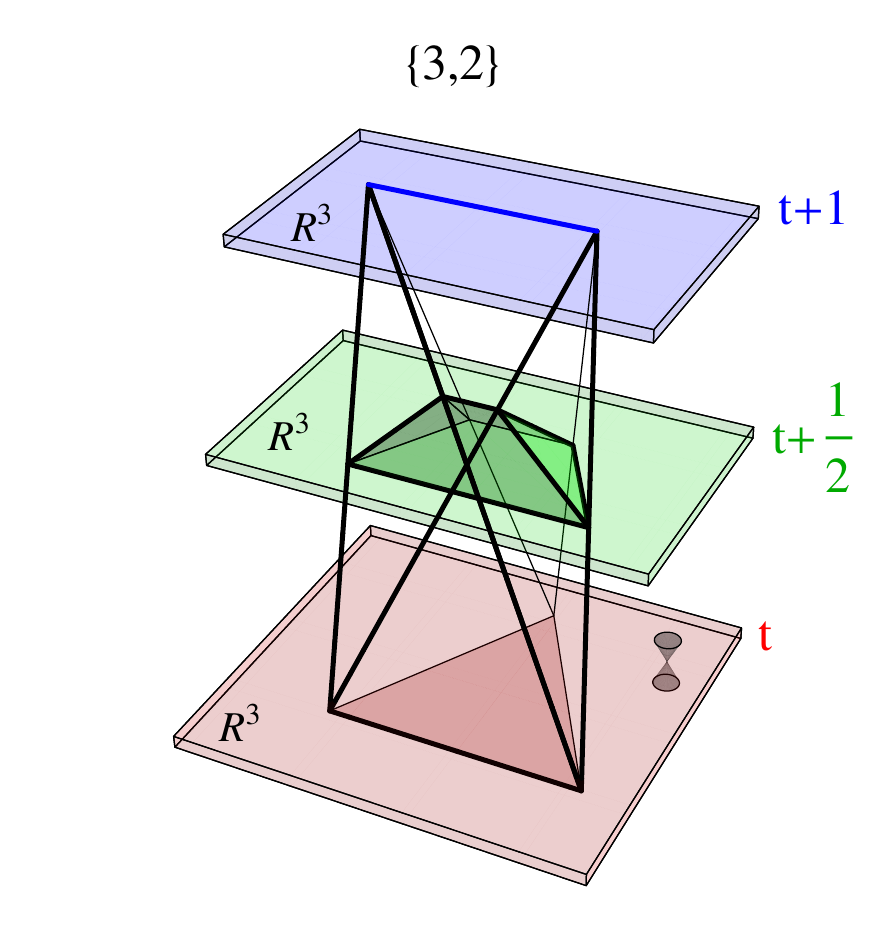}
	\caption{\small Visualization of fundamental building blocks in four-dimensional
		CDT. A $(4,1)$ simplex (left) has four vertices at (discrete) proper time $t$
		(forming a tetrahedron) and one vertex at time $t+1$. A $(3,2)$ simplex (right)
		has three vertices at time $t$ and two at $t+1$. Cauchy surfaces for non-integer $t$
		are built from a combination of tetrahedra (obtained by slicing $(4,1)$ simplices with
		hyperplanes of constant time) and triangular prisms (from $(3,2)$ simplices).}
	\label{Fig:simplex}
\end{figure}

In CDT the lattice is constructed from d-dimensional simplices glued together to form a {\it triangulation} which we view as  being a piecewise linear manifold. The geometry 
is fixed if we know the edge lengths of the simplices. If we for a moment assume 
that the edge lengths are all identical, this  length $a$ provides a UV cut-off which  
can in principle be removed by taking $a\to 0$. When such a limit is combined with 
keeping the physical observables fixed one will  approach the UV fixed point (if it exists). 
We refer to \cite{CDTRG} for a detailed discussion of this. The important and distinguishing feature of CDT compared to DT is the introduction of a well defined  causal structure compatible with global hyperbolicity. Each globally hyperbolic spacetime can be topologically foliated into  Cauchy spatial hypersurfaces 
$\Sigma$ of equal global proper time $T$ and the spacetime 
can be written as a product $\cM = \Sigma \times T$.   
In CDT one introduces such a foliation  by definition and any triangulation is 
then topologically $\cT = \Sigma \times T$. 
In $d=4$ dimensions any  such triangulation $\cT$ can be constructed from two kinds of 4-simplices. The $(4,1)$ simplex has 4 vertices on a spatial hypersurface in (integer) time $t$ and 1 vertex in $t\pm1$, and the $(3,2)$ simplex has 3 vertices in $t$ and 2 vertices in $t\pm 1$.  
Each spatial layer of equal integer time $t$ is by construction formed from equilateral tetrahedra being  parts of $(4,1)$ simplices. These tetrahedra are glued  together in such a way that the layer has (the chosen) topology $\Sigma$. 
The 4-simplices interpolate between consecutive spatial layers in such a way that the spatial topology $\Sigma$ as well as global topology   $\Sigma \times T$ is conserved for any (also non-integer) $t$. Therefore one can also distinguish spatial hypersurfaces of non-integer $t$, formed by slicing  4-simplices with three-dimensional hyperplanes of constant $t$. Such Cauchy surfaces are built from a combination of
tetrahedra, obtained by slicing $(4,1)$ simplices, and triangular prisms, from
$(3, 2)$ simplices, see Fig.\ \ref{Fig:simplex}. These building blocks are again glued
together, and by construction form a slice topologically isomorphic to $\Sigma$.
The formal path integral \rf{Zcont} is then defined by a sum over all triangulations $\cT$ obeying these topological constraints  
\beql{ZdiscrL}
\cZ = \sum_{\cT}e^{iS_R[\cT]}\ .
\eeq
{In Eq. \rf{ZdiscrL} $S_R$ is the discretized Hilbert-Einstein action obtained  following Regge's method for describing piecewise linear geometries \cite{regge} which can be expressed as a linear combination of $ N_{(4,1)}$,  $ N_{(3,2)}$ and $N_0$ denoting the total number of $(4,1)$ simplices, $(3,2)$ simplices and vertices, respectively. The dimensionless coupling constants  depend on the bare Newton's constant $G$, cosmological constant $\Lambda$ and the parameter $\alpha$, which defines asymmetry between lengths of time-like and space-like links in the lattice}:
\beql{alpha}
a_t^2 = - \alpha\  a_s^2  \quad ,  \quad \alpha > 0 \ .
\eeq    

In order to study the regularized path integral \rf{ZdiscrL} in $d=4$ dimensions one is forced to use numerical Monte Carlo techniques, which require changing from a quantum field theory to a statistical field theory regime. This method is commonly used in other lattice approaches as well, e.g. in lattice QCD, and it amounts to a Wick rotation from Lorentzian to Euclidean spacetime signature. In CDT  a kind of ``Wick rotation'' can be  defined due to the  imposed global proper time foliation. Technically it is done by analytically continuing  the asymmetry parameter $\alpha \to - \alpha \ (| \alpha | > 7/12)$ in the lower complex  $\alpha$ plane.
The rotation from positive to negative $\alpha$ values changes time-like links
into space-like links which is consistent with
$t_L \to t_E = i t_L$,
where $t_L$ is the real (Lorentzian) time and $t_E$ is the imaginary (Euclidean) time. The condition 
$| \alpha | >  7/12$ additionally ensures that all triangle inequalities
are fulfilled  in the Euclidean regime, which means that all simplices
become  parts of the Euclidean 4-dim space with well
defined positive volumes. Consequently the path integral \rf{ZdiscrL} becomes a partition function which can be studied numerically:
\beql{Zdiscr}
\cZ = \sum_{\cT}e^{iS_R^{(L)}[\cT]}\quad \to \quad  \cZ = \sum_{\cT}e^{-S_R^{(E)}[\cT] }.
\eeq
{The explicit parametrization of the action in the Euclidean case is
\beql{SRegge}
S_{R}^{(E)}=-\left(\kappa_{0}+6\Delta\right)N_{0}+\kappa_{4}\left(N_{(4,1)}+N_{(3,2)}\right)+\Delta \ N_{(4,1)} \  .
\eeq} 
Of course some properties of such a theory may and in general will   depend on the  spatial topology $\Sigma$ chosen. The choice of topology might  have 
an impact on semiclassical, or infrared,  properties of CDT as various classical solutions are consistent with various spacetime topologies and some solutions may be (dis)favoured or even not allowed in some spacetime topologies.
Most of the previous studies in 3+1 dimensions were done for spherical spatial topology, 
$\Sigma = S^3$, and one also introduced time periodic boundary conditions,
 resulting in the global spacetime topology $\cT = S^3 \times S^1$. This choice led to many interesting results,  including the identification of 4 distinct phases of spacetime geometry, called {$A$, $B$, $C_{dS}$ and  $C_{b}$}, respectively (for the most recent phase diagram see e.g. \cite{nilas}). 
 In phase  {$C_{dS}$} the dynamically generated semiclassical background geometry
 could be intepreted as a (Euclidean) de Sitter solution to  general relativity  and viewed 
 as the infrared limit  of CDT.  
 At the same time quantum fluctuations of the spatial volume 
 in phase {$C_{dS}$} were very well described by the standard minisuperspace action 
  obtained for the maximally symmetric geometry, i.e. for a  spatially isotropic and 
 homogenous spherical metric (see e.g. {\cite{physrep}} for a discussion).

\begin{figure}[t]
\begin{center}
\includegraphics[width=0.49\textwidth]{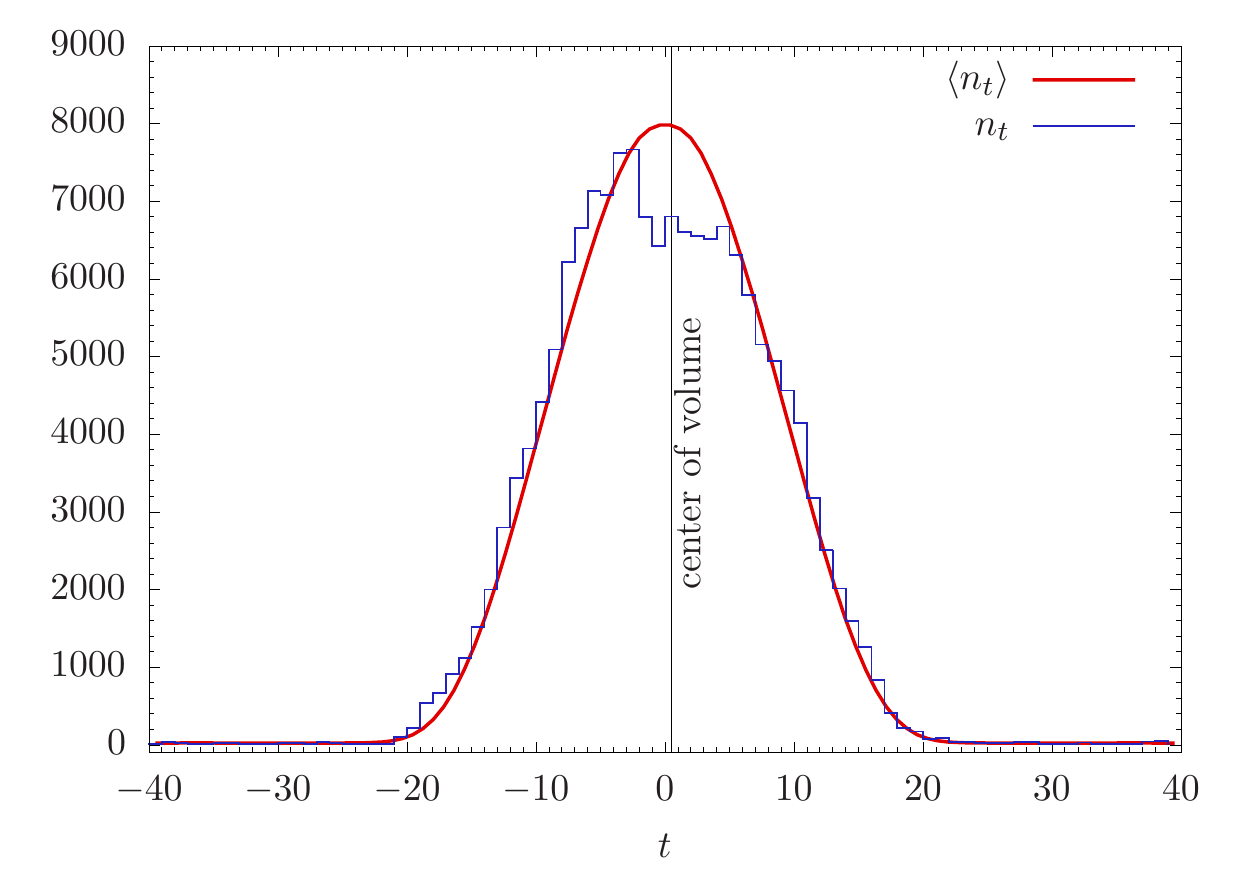}
\includegraphics[width=0.49\textwidth]{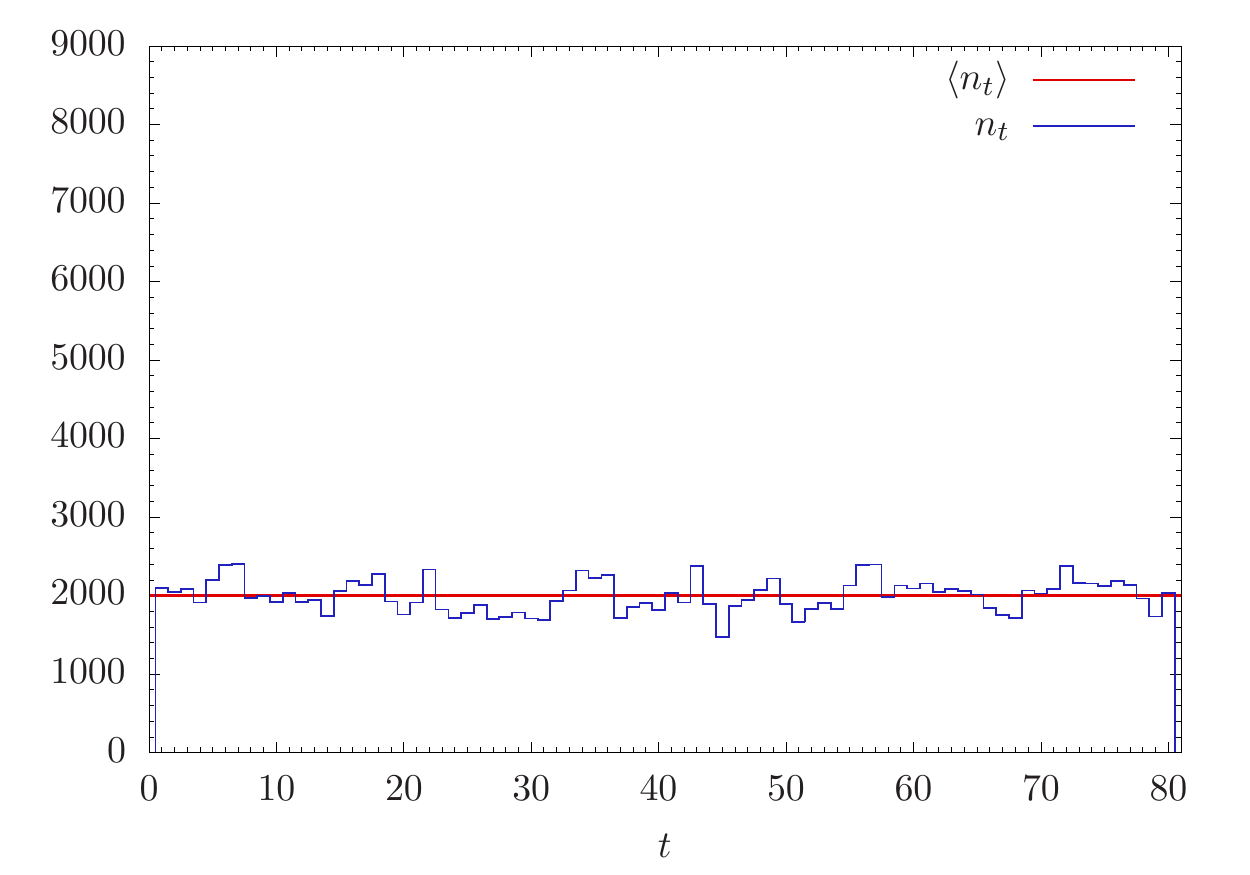}
\end{center}
\caption{
The average (red) and typical (blue) volume profiles for spherical spatial topology
({left} figure) and for toroidal  spatial topology ({right} figure) in the so-called
de Sitter phase {($C_{dS}$)}. In both cases the time direction has length 80 (in lattice units) with 
periodic boundary conditions. In the case of spherical spatial topology the ``effective''
time extent is approximately 40 lattice units and scales as $N^{1/4}$, $N$ being the total
four-volume of the universe, while $n_t$ scales as $N^{3/4}$. This is in contrast to 
the case of  toroidal  spatial topology 
where $n_t$ is macroscopic for all $t$ and scales as  $N/T$, $T$ being the time extent.}
\label{fig:spheretorus}
\end{figure}

We want to investigate to what extent these results in phase {$C_{dS}$} remain valid when we 
change the spatial topology from $S^3$ to  $\Sigma =  T^3 \equiv S^1 \times S^1 \times S^1$.
In \cite{torus}  we showed that this change in topology resulted in a dramatic 
change in the dynamically generated background geometry, {as illustrated 
in Fig.\ \ref{fig:spheretorus}},  but that this 
background geometry was still well described by a suitable toroidal minisuperspace
action. The purpose of the current work is to corroborate the results reported in \cite{torus}
and analyze the quantum fluctuations around the new semiclassical geometry.

\section{The covariance matrix}
\label{sec:cov}

As described above the setup is the following: we have a foliation, labeled by 
a (discrete) time variable $t$. Associated with each $t$ we have a 
spatial volume $n_t$, which is the number of tetrahedra (by convention multiplied 
by a factor of 2 to give the number of four-simplices which have four 
vertices at time-slice $t$) used to 
construct space at time $t$. The time $t$ takes integer values in a range $[1,T]$. 
The Monte Carlo simulations will create  distributions of $n_t$'s for each $t$, and the 
average values $\la n_t\ra$ will be our spatial volume profiles. By measuring correlations
\beq\label{j1}
 C_{i j} = \langle \eta_i \eta_j \rangle = \left\langle (n_i - \langle n_i \rangle) (n_j - \langle n_j \rangle) \right\rangle,
 \eeq
where {$\eta_t = n_t - \bar n_t$ is the deviation of the spatial volume $n_t$ 
from the mean value $\bar n_t = \left\langle n_t \right\rangle$}, we can  
reconstruct a semiclassical  effective 
action of $n_t$ which reproduces $C_{ij}$. In a suitable minisuperspace parametrization
the spatial volume (i.e.\ here $n_t$) is proportional to {$a^3(t)$, where $a(t)$} denotes
the scale factor of the universe. Thus constructing an effective action $S[n_t]$ amounts
to constructing a minisuperspace action, but with the important twist that we 
obtain this action by actually integrating out other degrees of freedom than the 
scale factor, rather than putting in by hand a special form of geometry only 
depending on one dynamical variable, the scale factor {$a(t)$}. $C_{ij}$ is denoted 
the covariance matrix. It is related to  the effective action
to quadratic order in the fluctuations  as follows:
\beq\label{j2} 
{S[n = \bar n + \eta] = S[\bar n] + \frac{1}{2} \eta_t P_{t t'} \eta_t' + O(\eta^3),}
 \eeq
where $P$ is the inverse of the covariance matrix,
\beq\label{j3}
 P_{i j} = [C^{-1}]_{i j} = \left. \frac{\partial^2 S[\{n_t\}]}{\partial n_i \partial n_j}\right|_{{n =\bar n }} . 
 \eeq

The effective action has a kinetic term which in a first approximation 
depends of $n_t -n_{t\pm 1}$ and a potential term which depends on $n_t$.
The covariance matrix method for reconstructing the effective action 
worked well for a spherical topology \cite{transfermatrix} where the  average 
spatial volume $\la n_t \ra$ was a function of $t$ 
($\la n_t \ra \propto  N^{3/4} \cos^3 (c \cdot t/N^{1/4})$ where $N$ denotes the total spacetime 
volume, i.e.\ the number of four-simplices and $c$ is a constant).
However, in the toroidal case with time-periodic boundary conditions, 
one has to a first approximation $\la n_t \ra \propto N/T$, a situation which is 
not optimal if we want to determine the kinetic term as a function of $n_t -n_{t\pm 1}$.
\begin{figure}
\begin{center}
\includegraphics[width=0.9\textwidth]{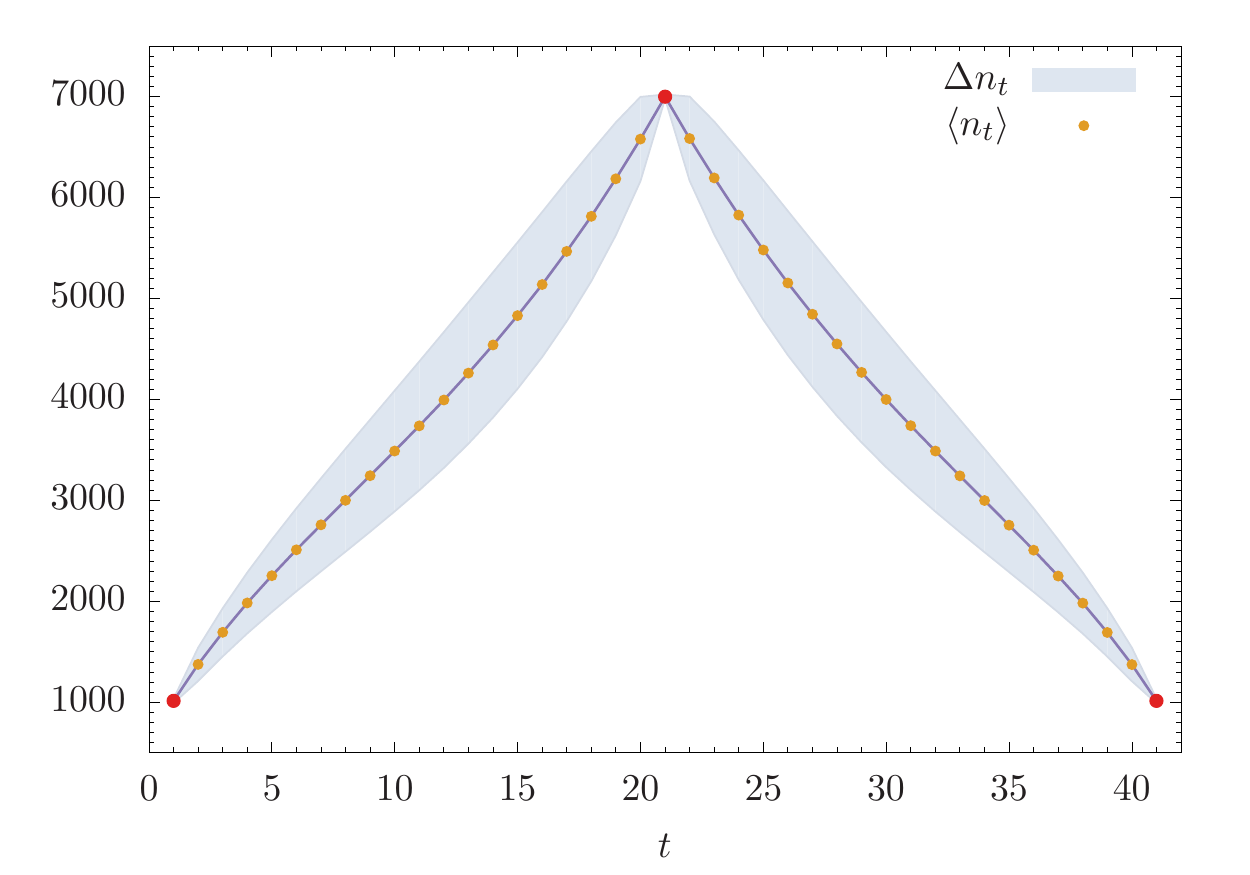}	
\caption{Average volume profile $\langle n_t \rangle$
measured for bare coupling constants $\kappa_0 = 2.2, \kappa_4 = 0.9225, \Delta = 0.6$
and time period $T = 40$. 
Red dots denote time slices with fixed volume,
$\hat{{n}}_1 = 1000$ and $\hat{{n}}_h = 7000, h = T / 2 + 1$.}
\label{fig:average}
\end{center}
\end{figure}
We have chosen to circumvent this problem by enforcing the volume profile to span some range. One could introduce boundary conditions in  time direction which are incompatible 
with a constant $\langle n_t \rangle$ solution. However, it is simpler to add to the bare action \rf{SRegge}
used in the Monte Carlo simulations a correction $S_{\mathrm{fix}}$,
\begin{equation}\label{eq:sfix}
S_{\mathrm{fix}} = \frac{1}{2} \varepsilon \left[ (n_1 -\hat{{n}}_1)^2 
+ (n_h - \hat{{n}}_h)^2 \right],
\end{equation}
which fixes volumes of two slices, namely $n_1$ and $n_h$.
It is convenient to choose $h = T / 2 + 1$,
so that the volumes of the first slice $n_1$ and
a middle slice $n_h$ appear in the correction.
Because of the time periodic boundary condition
slices $t = 1$ and {$t = T+1$} are identified and the 
time separation between $n_1$ and $n_h$ is the same as between
$n_h$ and $n_{T + 1} = n_1$
which ensures symmetry $n_t \leftrightarrow n_{T + 2 - t}$ up to periodicity.
The volumes $n_1$ and $n_h$ are not rigidly fixed, but fluctuate around 
the prescribed values $\hat{{n}}_1$ and $\hat{{n}}_h$, respectively,
with an amplitude dependent on $\varepsilon$.
Finally, the effect of $S_{\mathrm{fix}}$ can be easily removed from the final results
as will be discussed below.
{The advantage of this method is that it allows for 
a measurement of the effective action for a number of spatial volumes 
$\bar{n}_t$
in a single simulation.
Thus it simplifies the analysis and reduces statistical errors.}

All results and measurements presented in this work were performed
for a specific choice of bare coupling constants which are supposed 
to lie in the analog of the de Sitter phase (phase {$C_{dS}$}) observed for the spherical topology. 
Figure \ref{fig:average} shows an average volume profile,
\beq\label{j4}
{\bar n_t} = \langle n_t \rangle, \quad t = 1, \dots, T 
\eeq
for $T = 40$ time slices.
Points corresponding to the two slices with fixed volume, $n_1$ and $n_h$, 
are marked with red dots. The average volume profile spans a range 
between $\hat{{n}}_1 = 1000$ and $\hat{{n}}_{21} = 7000$.
The light blue envelope illustrates the amplitude of volume fluctuations $\Delta n_t$.
Because the action does not include a total volume fixing term
the volume profile is very sensitive to the choice of the bare cosmological 
coupling constant $\kappa_4$.
To obtain a desired average total four-volume the coupling constant 
$\kappa_4$ has to be precisely fine-tuned. While this is a slight complication 
in the numerical simulations compared to {adding by hand a} volume fixing term
(as was done in the simulations where the spatial topology was that of $S^3$),
the absence of the total volume fixing term simplifies the analysis of the measured 
covariance matrix (for complications created by a volume fixing term see
e.g. \cite{physrep}.)
\begin{figure}
\begin{center}
\includegraphics[width=0.6\textwidth]{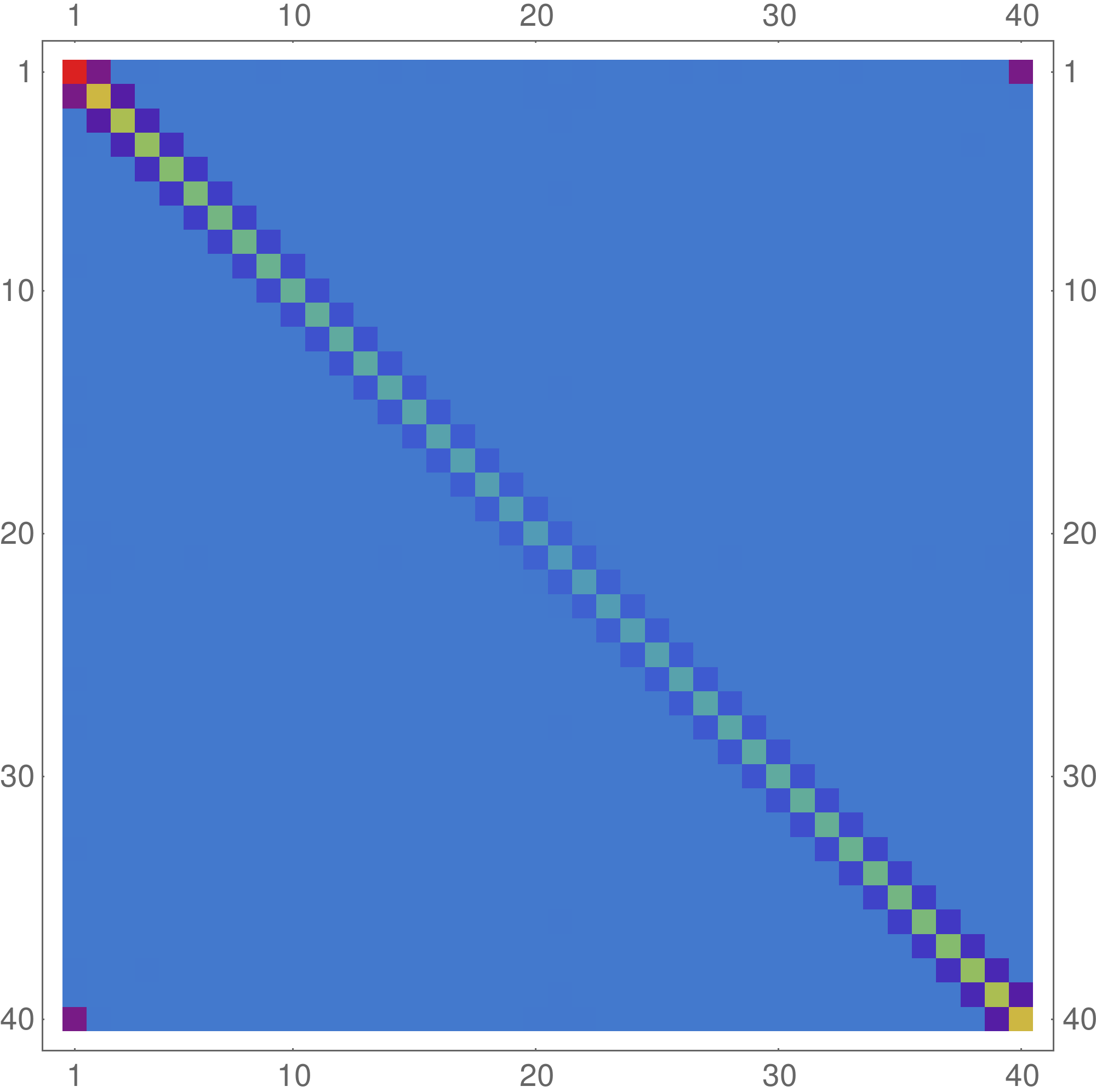}
\caption{Density plot of the covariance matrix inverse, $P = C^{-1}$.
The tri-diagonal structure is well visible.
The entries of $P$ outside the tri-diagonal are zero up to  numerical noise.}
\label{fig:pmatrix}
\end{center}
\end{figure}
As mentioned above, it is easy to correct for the addition of $S_{\mathrm{fix}}$ defined in  (\ref{eq:sfix})
{ to the bare CDT action (\ref{SRegge})}. $S_{\mathrm{fix}}$ is quadratic and by \rf{j3} {we only have} to subtract 
$\varepsilon$ from matrix elements $P_{1 1}$ and $P_{h h}$.
The {time} reflection symmetry of the system, $n_t \leftrightarrow n_{T + 2 - t}$,
can also lead to a reduced measurement uncertainty by applying a symmetrization procedure
to the average volume profile and covariance matrix,
\begin{align*}
\langle n_t \rangle &\leftarrow \frac{1}{2}\left(\langle n_t \rangle + \langle n_{\tilde{t}} \rangle\right), \\
C_{i j} &\leftarrow \frac{1}{2} \left( C_{i j} + C_{\tilde{i} \tilde{j}}   \right) 
	+ \frac{1}{4} (\langle n_i \rangle - \langle n_{\tilde{i}})(\langle n_j \rangle - \langle n_{\tilde{j}} \rangle ),
\end{align*}
where $\tilde{t} = T + 2 - t$.
Such operation is equivalent to doubling  the {measurement statistics}.

A density plot of the inverse covariance matrix $P$ is shown in Fig.\ \ref{fig:pmatrix}.
The tridiagonal form is clear and suggests that the effective action describing fluctuations of $n_t$ is quasi-local in time,
\begin{equation}\label{eq:sku}
S[\{n_t\}] = \sum_t \big( K(n_t, n_{t+1}) + U(n_t) \big).
\end{equation}
The function $K$ describes the kinetic part of the effective action.
It couples volumes of adjacent slices and provides  non-zero subdiagonal elements of $P$. 
Since both $C$ and $P$ are symmetric matrices, the function $K(n, m)$ has to be symmetric in its arguments, $K(n, m) = K(m, n)$.
The potential part, described by the function $U(n)$, contributes only to the diagonal.

\subsection{The kinetic term}
\label{sec:cov:kin}

\begin{figure}
\includegraphics[width=0.9\textwidth]{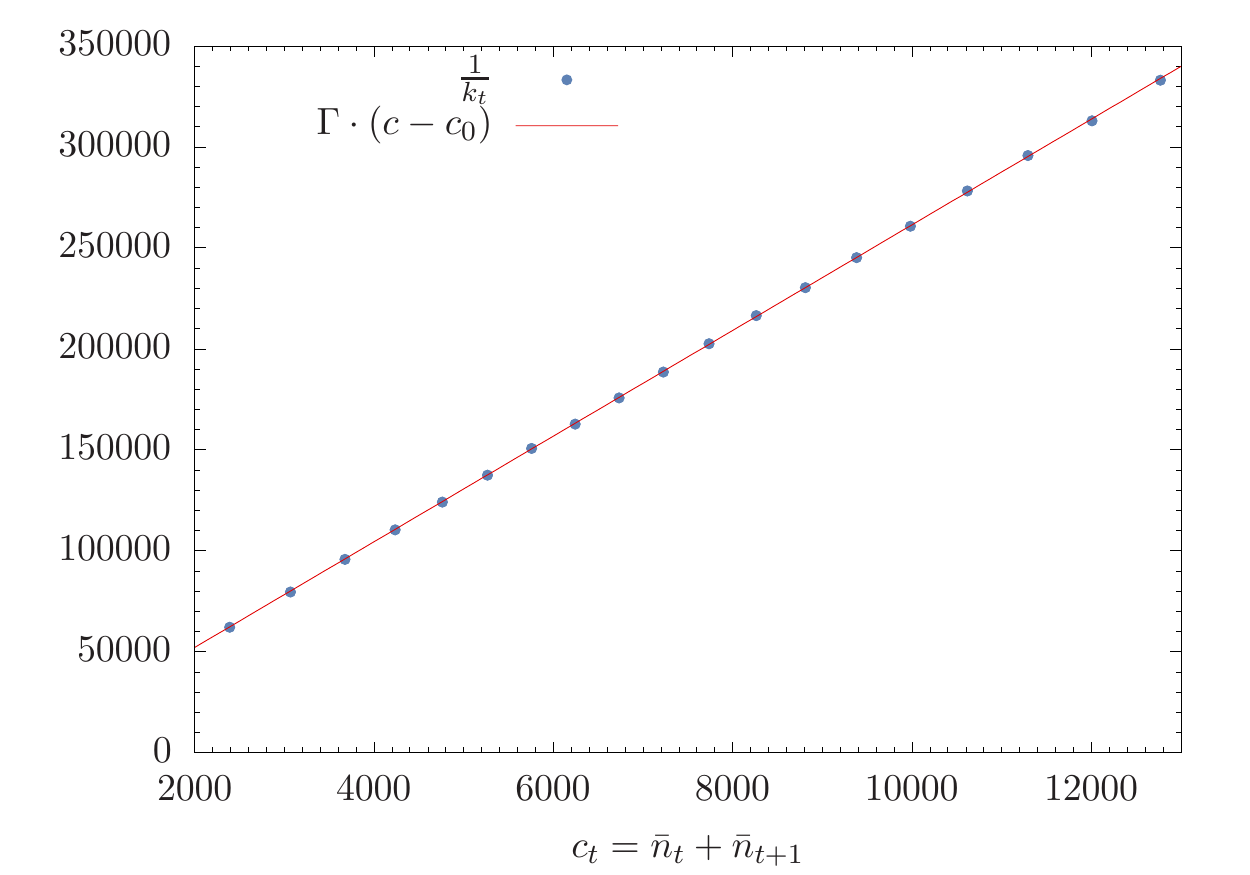}
\caption{Inverse of the kinetic term $k_t = - \frac{1}{2} P_{t\,t+1}$ as a function of volume $c_t = \langle n_{t} \rangle + \langle n_{t + 1} \rangle$.
A linear fit $\Gamma \cdot (c - c_0)$ is very accurate.}
\label{fig:kincov}
\end{figure}

The kinetic part of the effective action makes, from a numerical point of view, 
a dominating contribution to the path integral, 
and thus it is measured with the smallest uncertainty.
Because  subdiagonal elements of the matrix $P$ depend only on the kinetic term, {while diagonal elements depend both on the kinetic and the potential term,
the kinetic part has  to} be determined before the potential term.
To extract the kinetic term of the effective action we proceed  as in {\cite{semiclassical}}.
In Sec.\ \ref{sec:tm} we describe an alternative method which uses the  transfer matrix,
and which turns out to be even more accurate.

A semiclassical expansion of the action (\ref{eq:sku}),
\beq\label{j5}
{ S[n = \bar n + \eta] = S[\bar n] +} \sum_t k_t \cdot (\eta_{t+1} - \eta_t)^2 + u_t \eta_t^2 + O(\eta^3),
 \eeq
introduces the kinetic coefficients $k_t$ and potential coefficients $u_t$.
The kinetic coefficient is equal to the second derivative of the kinetic term $K$,
present in (\ref{eq:sku}),
and to the subdiagonal elements of matrix $P$,
\begin{equation}\label{eq:pttkt}
{P_{t\,t+1} = - 2 k_t = \left. \frac{\partial^2 K(n, m)}{\partial n \partial m}\right|_{n = \bar n_t, m = \bar n_{t+1}} }.
\end{equation}
Motivated by the results for the {spherical spatial topology \cite{physrep} and initial results for the toroidal topology \cite{torus}}, 
we expect $\frac{1}{k_t}$ to be a linear function of {$c_t = \bar n_t + \bar n_{t+1}$}.
This is indeed supported by the results which are presented in Fig.\ \ref{fig:kincov}.
The blue dots {are $\frac{1}{k_t}$ coefficients} extracted from the inverse covariance matrix (\ref{eq:pttkt}),
the red line is a linear fit $\Gamma \cdot (c - c_0)$ with $\Gamma = 26.2 \pm 0.1,\, c_0 = 5.0 \pm 0.1$.
The result is not only in agreement with previous results {for toroidal} topology \cite{torus},
but the coefficient $\Gamma$ is also very close to the value obtained for {the spherical topology} \cite{physrep}.

Expanding the right-hand side of equation (\ref{eq:pttkt}) with respect to $\frac{n - m}{n + m}$,
the result {
\[ k_t = \frac{1}{\Gamma} \frac{1}{\bar n_t + \bar n_{t + 1}} \]
becomes the leading-order term for the kinetic part
\[ K(n, m) = \frac{1}{\Gamma} \frac{(n - m)^2}{n + m} . \]
}The shift $c_0$ can be  neglected since its value  is small.
For a discussion of higher order corrections to the kinetic 
term we refer to \cite{semiclassical}.

\subsection{The potential term}\label{sec:cov:pot}

\begin{figure}
\begin{center}
\includegraphics[width=0.9\textwidth]{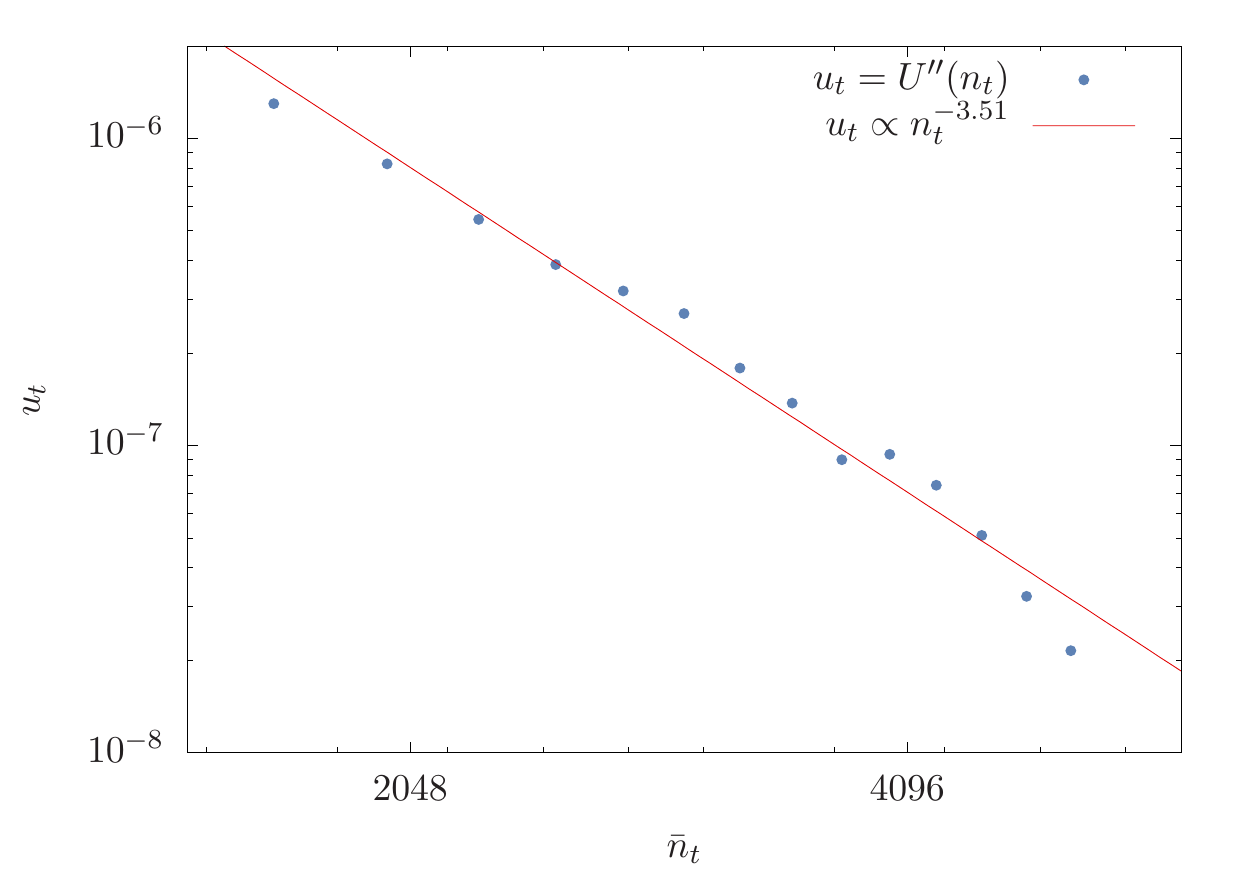}								
\caption{Log-log plot of the potential { coefficients} $u_t$ as a function of $\langle n_{t} \rangle$.}
\label{fig:potcov}
\end{center}
\end{figure}

The diagonal of the inverse covariance matrix $P$ is affected both by the kinetic term and the potential term of the effective action,
\[ P_{t\,t} = k_t + k_{t - 1} + u_t = \left. \frac{\partial^2 S[\{n_t\}]}{\partial n_t^2 }\right|_{n_t = {\bar n_t}} . \]
Figure\ \ref{fig:potcov} plots the potential coefficients $u_t$ against $\bar n_t$ in a log-log scale,
where $u_t$ was determined by {subtracting from the measured data $P_{tt}$ the  already extracted values of $k_t$}, as 
described in the previous Section.
The results show a power-law behavior,
\[ u_t = \const \times \langle n_{t} \rangle^{\gamma - 2}, \quad \gamma \approx -1.5. \]
Since { $u_t = \left. \frac{\partial^2 U(n)}{\partial n^2 }\right|_{n = \bar n_t}$,
the potential term can be written as 
\[ U(n) = \mu \, n^{\gamma} + \lambda \, n. \]}

\subsection{Derivative of the potential term}

\begin{figure}
\includegraphics[width=0.9\textwidth]{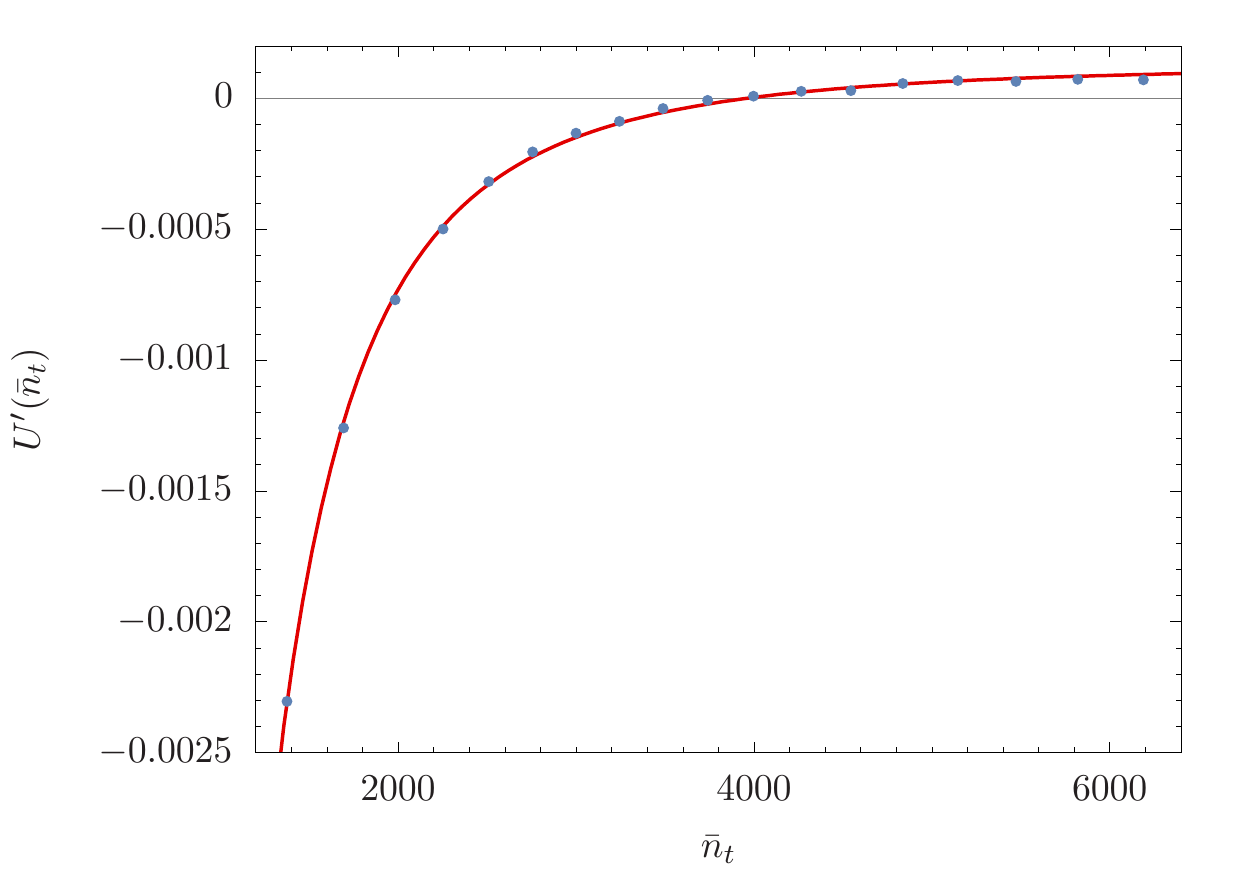}
\caption{First derivative of the potential extracted from the classical trajectory (blue dots) 
and a power-law fit {$U'(n) = \lambda + \const \times n^{\gamma -1}$},{ $\gamma \approx -1.78 \pm 0.1$}. }
\label{fig:uprim}
\end{figure}

The principle of least action states that  the classical trajectory ${\bar n_t} = \langle n_t \rangle$
is an extremum of  the action
\beq\label{j6} 
{S[\{n_t\}] = \sum_t \big(K(n_t, n_{t+1}) + U(n_t)\big)}. 
\eeq
{Consequently}  the first derivative of the action has to vanish, 
$\left.\frac{\delta S}{\delta n_t}\right|_{n_t = {\bar n_t}} = 0$ and { thus} knowing ${\langle n_t \rangle}$
the derivative of the potential can be determined from the kinetic term:
\begin{equation}\label{eq:uprimk}
{U'(\bar n_t) = -\left[ \frac{\partial K(n_{t+1}, n_{t})}{\partial n_t} + \frac{\partial K(n_{t}, n_{t - 1})}{\partial n_t} \right]_{n_t = \bar n_t}}.
\end{equation}
The first derivative of the potential can thus be extracted from the measured 
classical trajectory ${\bar n_t} = \langle n_t \rangle$
, using the 
presumed form of the kinetic term   obtained in Sec.\ \ref{sec:cov:kin}, namely
\[ K(n, m) = \frac{(n - m)^2}{\Gamma (n + m - c_0)} . \]
The result is shown in Fig.\ \ref{fig:uprim}.
A power-law function fits well to the data,
with the exponent {$\gamma \approx -1.78 \pm 0.1$}. The value is  slightly larger 
than the one  obtained using  the covariance matrix method.

\subsection{The effective action}

Summarizing the results, the full discrete effective action,
determined via the covariance matrix within the semiclassical approximation,
is given by
\begin{equation}\label{eq:scov}
S[\{n_t\}] = \sum_t \left[ \frac{1}{\Gamma} \frac{(n_{t + 1} - n_t)^2}{n_{t + 1} + n_t} + \mu \, n_t^{\gamma} + \lambda \, n_t \right].
\end{equation}
The outcome, up to a slight difference in the value of {$\gamma\approx -1.5$}, is in agreement with 
\cite{torus}{, where the similar effective action was extracted using a collection of (inverse) covariance matrices $P$ measured for various lattice volumes $N$ and time periods $T$ for a system with no imposed boundary volume fixing  terms, i.e. with constant volume profiles $\langle n_t \rangle \propto N/T$.} The continuous counterpart of the discrete action (\ref{eq:scov}) is
\begin{equation}\label{eq:scovcont}
S[v] = \int \dd t \, \left[ \frac{1}{\tilde\Gamma} \frac{\dot{v}^2}{v} + \tilde\mu \, v^{\gamma} + \tilde\lambda \, v \right], \quad \gamma \approx -1.5
\end{equation}
where $v(t)$ is the physical spatial volume.

{In the above one recovers the kinetic term which can be also obtained by a suitable minisuperspace reduction of the usual Einstein-Hilbert action, i.e. by requiring the spatially homogenous and isotropic metric and freezing all degrees of freedom but the scale factor. However it is  important to note that the nature of the effective action (\ref{eq:scovcont})  is very different from the usual minisuperspace action as it was obtained from a full lattice model  after integrating out (averaging over) all other degrees of freedom but the scale factor, rather than freezing them. Although the kinetic term in  (\ref{eq:scovcont}) superficially resembles the standard minisuperspace kinetic term it also has opposite sign. It is possible due to a very subtle interplay between the bare Regge-Hilbert-Einstein action of CDT quantum gravity (\ref{SRegge}) and entropy of configurations which play  equally important roles in the path integral (\ref{Zdiscr}). It turns out that in CDT formulation the entropy factor leads to the same kind of the effective action as the minisuperspace reduction but with oposite sign and it dynamically corrects the wrong sign of the usual minisuperspace action.\footnote{{Standard Euclidean minisuperspace action is unbounded from below due to  negative sign of the kinetic term. The sign of the action does not matter for a classical trajectory but in the path integral formalism the unbounded action causes the path integral to be completely dominated by arbitrarily large fluctuations of the conformal mode.}}
 Analytic calculations which support this picture can be found in 
\cite{loll}. The potential part of the effective action (\ref{eq:scovcont}) does not have its classical counterpart, as no such a term is present in the standard minisuperspace reduction of  toroidal geometry, and thus it can be treated as a pure quantum correction.
} 

\section{The classical trajectory}\label{sec:classical}	

\begin{figure}[t]
\centering
\subfigure[Torus.]{\label{fig:torus_potential} \includegraphics[scale=0.45]{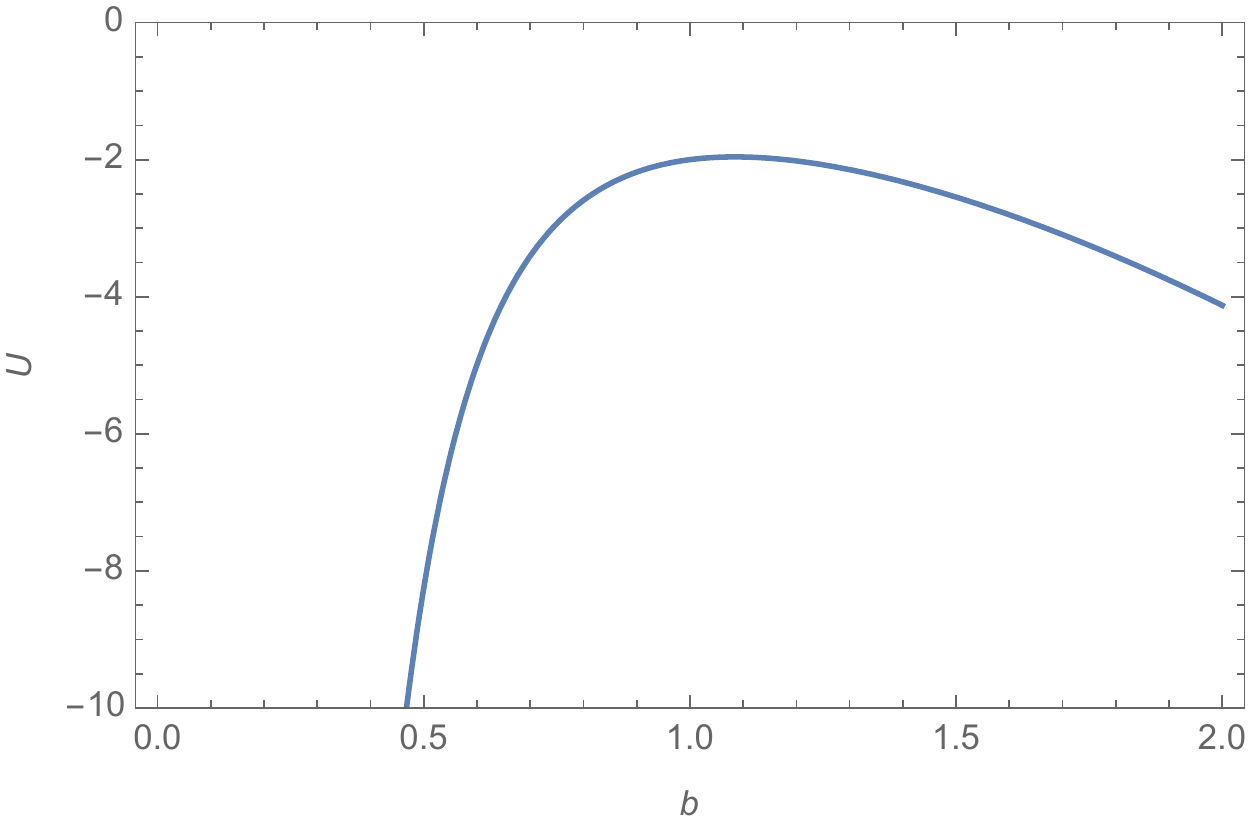}}
\subfigure[Sphere.]{\label{fig:sphere_potential} \includegraphics[scale=0.45]{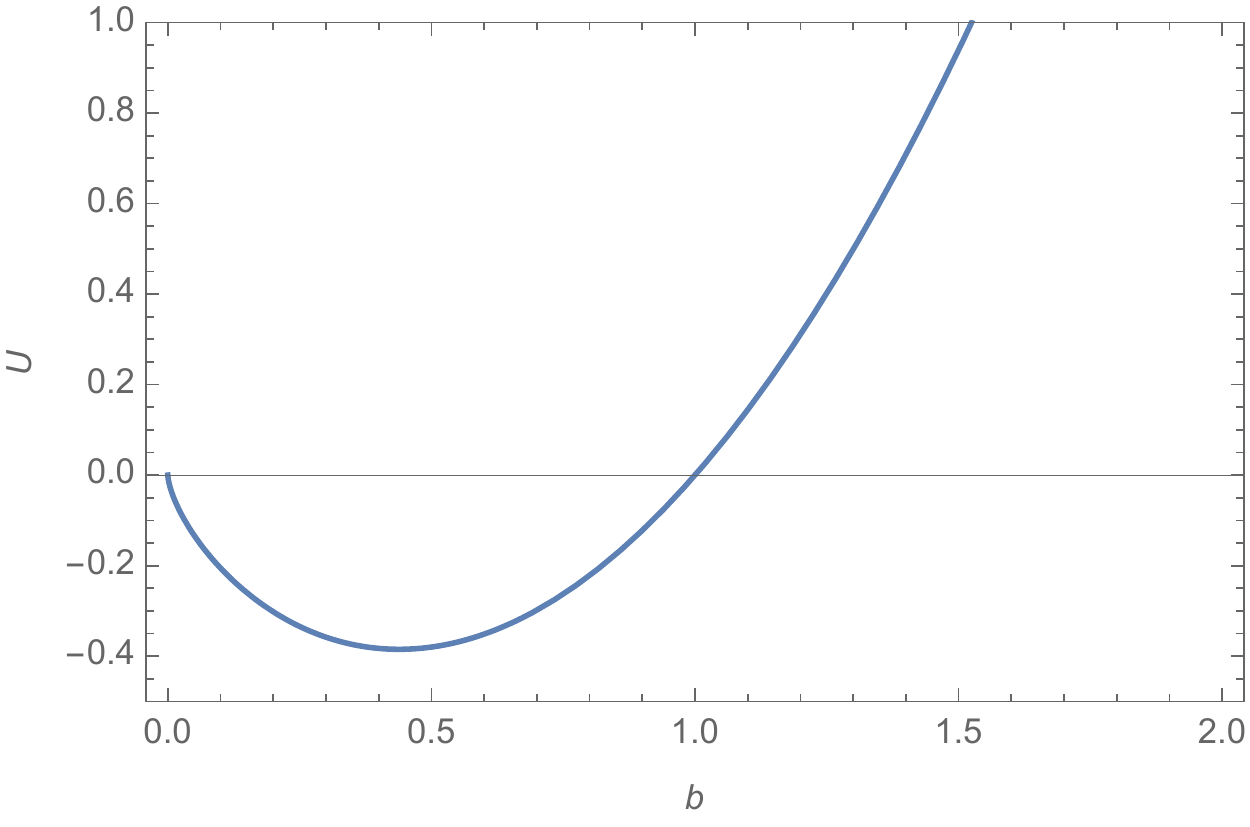}}
\caption{The effective potential $U[b]$ for the sphere and torus cases.}
\label{fig:potentials}
\end{figure}

\begin{figure}[t]
\begin{center}
    \includegraphics[scale=0.6]{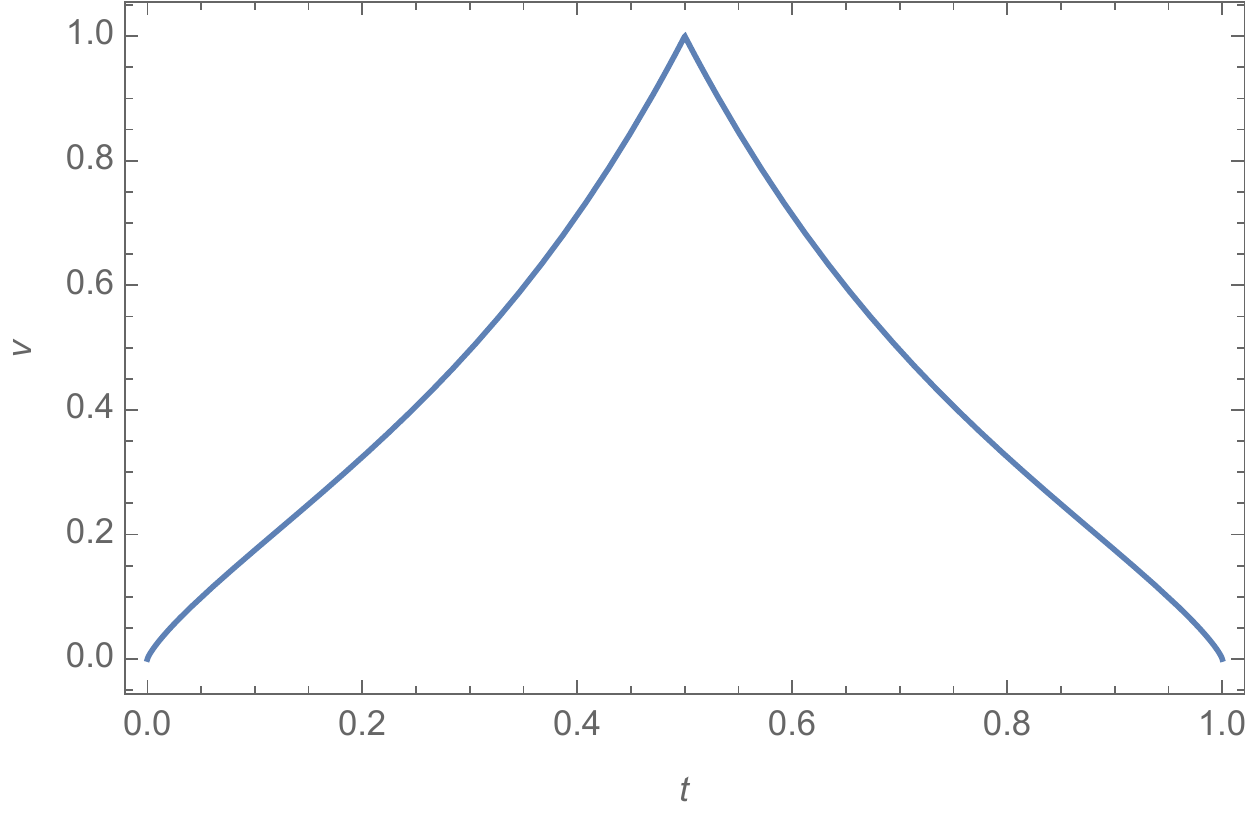}
\end{center}
\vspace{-0.7cm}
\caption{The classical trajectory interpolating between zero initial volume and some intermediate volume.}
\label{fig:hersheys_kiss}
\end{figure}

In this section we study the equations of motion corresponding to the {effective} 
action  \rf{eq:scovcont}, which we determined using the covariance matrix.
$v(t)$ denoted the spatial volume at time $t$, but for the purpose of 
analyzing the equations of motion we will talk about $v(t)$ as the position of a "particle". 
Thus our starting point is the particle action
\begin{equation}
S = \int \mathrm{d}t\  \left[ \frac{\dot{v}^2}{\Gamma v} + \mu v^\gamma + \lambda v \right], \label{eq:action}
\end{equation}
where $\gamma = -3/2$.
To simplify the problem of finding a trajectory which minimizes action (\ref{eq:action})
we introduce a change of variables
\[ v(t) = b^2(t). \]
The advantage of this substitution is that it produces an equation of motion with a simple physical interpretation as a particle moving in a particular potential. The Lagrangian for $b(t)$ is given by
\[ L[b] = \frac{4}{\Gamma} \dot{b}^2 + \mu b^{2 \gamma} + \lambda b^2 \]
and the corresponding equation of motion is
\[ Q[b] = \frac{\partial L}{\partial b} - \frac{\dd}{\dd t} \frac{\partial L}{\partial \dot{b}}
= - \frac{8}{\Gamma} \ddot{b} + 2 \mu \gamma b^{2 \gamma -1} + 2 \lambda b = 0 .\]
This describes the motion of a classical particle of mass {$\frac{8}{\Gamma}$} in the potential $U[b] = - \mu b^{2 \gamma} - \lambda b^2$ 
(with $\mu, \lambda > 0$). 
The equation of motion can be written as a total derivative of the total energy $W[b]$,
\begin{equation}\label{eq:QW}
Q[b] = - \frac{1}{\dot{b}} \frac{\dd}{\dd t} W[b] = 0 \ \Leftrightarrow \ 
W[b] = \frac{4}{\Gamma} \dot{b}^2 - \mu b^{2 \gamma} - \lambda b^2 = \const,
\end{equation}
where $W[b] = H = \const$ is just the law of conservation of total energy.
It has a formal solution
\[ \int \frac{2 \dd b}{\sqrt{\Gamma} \sqrt{H + \lambda b^2 + \mu b^{2 \gamma}}} = t - t_0, \]
where $H$ is the conserved total energy.
The above equation does not seem {analyticaly} solvable in the general case, even for $\gamma = -  \frac{3}{2}$.
For $H = 0$, the solution to equation (\ref{eq:QW}) is
\begin{equation} \label{eq:sinh}
v(t) = \left [ \sqrt{\frac{\mu}{\lambda}} \sinh [\alpha (t - t_0)]  \right ]^d, \quad d = \frac{2}{1 - \gamma}, \ \alpha =\frac{\sqrt{\lambda \Gamma}}{d}.
\end{equation}
However, even for $H \neq 0$, one can understand the classical trajectory fairly simply just from the form of the potential. The generic form of the potential in the {toroidal}  case is plotted in Fig.~\ref{fig:torus_potential}, in comparison to the {spherical} case in Fig.~\ref{fig:sphere_potential} (the axis scales are immaterial; only the shape of the potential is of interest).
Since $U[b]$ for the torus is strictly negative, $H=W[b] = 0$ implies that the {"particle"} always has some nonzero velocity. If the initial velocity is positive (towards increasing $b$), then the  {"particle"} will always roll towards ever-increasing $b$, or increasing volume, no matter the initial value of $b$. If the initial velocity is negative (towards decreasing $b$), then the  {"particle"} will always roll towards $b=0$, {or decreasing volume,} again no matter the initial value of $b$. In fact, this behavior persists even as we lower the value of $W[b]$ down to some lower critical value given by the maximum value of $U[b]$,
\[ 
U_{\text{max}} = - (1- \gamma ) ( - \gamma)^{\frac{\gamma}{1- \gamma}} \biggl( \frac{\mu}{\lambda^{\gamma}} \biggr)^{\frac{1}{1- \gamma}} \approx -2 \mu^{2/5} \lambda^{3/5} \approx -2.52,
\]
where we have used the values of $\mu \approx 2.86 \times 10^5$ and $\lambda \approx 3.5 \times 10^{-4}$ found in Section 4.2. We do not know the exact form of the solution except for $W[b] = 0$, but it will have the same qualitative behavior. For example, in Fig.~\ref{fig:average}, the initial volume {(at $t=0$)} is below the position of the maximum of the potential and the intermediate peak volume {(at $t=T/2+1$)} lies above. The system must have total energy above the maximum value of the potential in order for the classical trajectory to even reach the intermediate point. The volume increases quickly, slows down near the maximum of the potential and then increases quickly again as it rolls over the potential hump. We do not necessarily know the value of the total energy in Fig.~\ref{fig:average}, and so the trajectory may not be exactly given by the sinh function in \eqref{eq:sinh}, but the trajectory nevertheless exhibits the same qualitative behavior. For example, Fig.~\ref{fig:hersheys_kiss} shows the classical trajectory, where the initial rise takes the form of a growing $\sinh$ function and the subsequent fall a decaying $\sinh$ (the duration and amplitude are set to 1).

\section{The effective transfer matrix}
\label{sec:tm}

\begin{figure}
\begin{center}
\includegraphics[width=0.96\textwidth]{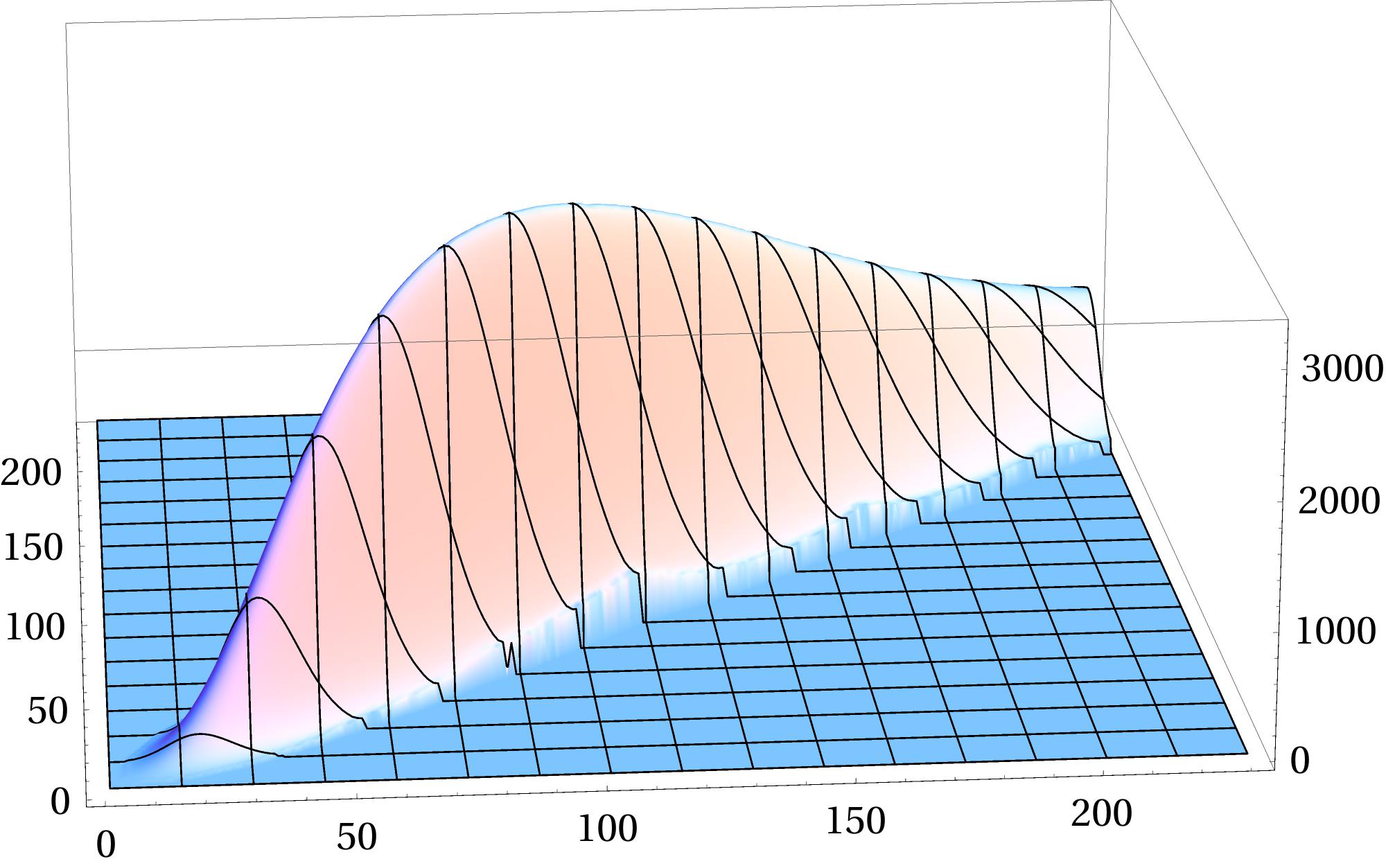}
\end{center}
\caption{Visualization of the effective transfer matrix. {Horizontal axes denote spatial volume $n$ and $m$, respectively, and vertical axis denote value of matrix elements  $\langle n| M | m \rangle$}.}
\label{fig:tmatrix}
\end{figure}

The so-called effective transfer matrix method allows us to measure 
directly the effective action.
It does not involve derivatives of the action, thus reducing numerical errors, 
and it allows us  to determine the value of $\lambda$.
For details, we refer the reader to \cite{transfermatrix}.

The effective transfer matrix, denoted as $M$, is directly related to the probability of encountering a configuration with given volume profile,
\beq\label{eq:efftm}
 P^{(T)} (n_1, \dots, n_T) \ {\propto} \ \langle n_1 | M | n_2 \rangle \langle n_2 | M | n_3 \rangle \dots  \langle n_T | M | n_1 \rangle.
 \eeq 
The probability can be measured in Monte Carlo simulations simply by 
counting the number of tetrahedra $n_t$ at each time slice $t$ for each 
independent four-dimensional triangulation generated during the simulation.
The setup is particularly easy for a  system with only $T = 2$ time slices {where, up to normalization,} there is a simple relation between the measured probability and the 
matrix elements $\la n | M |m \ra$:
\beq\label{j10}
\langle n | M | m\rangle =  \sqrt{P(n, m)} .
\eeq
However, it is also possible to use larger values of $T$ and extract 
the matrix elements by taking various ratios between probabilities,
see \cite{transfermatrix} for details. The results
reported here have used $T=2$. 

Due to the time-foliation present in the CDT formalism, CDT has {by definition} a transfer matrix 
providing us with the probability (amplitude) that the system evolves from a 
spatial geometry (i.e.\ three-dimensional triangulation $\cT^{(3)}_t$ of a spatial slice) 
at time $t$ to another spatial geometry at time $t+1$. This transfer matrix 
$\la \cT^{(3)}_{t+1} | \cM | \cT^{(3)}_t\ra$ 
is different from $\la n_{t+1}|M |n_t \ra$ defined in \rf{j10} 
which only depends on the spatial volumes $n_{t+1}$ of $\cT^{(3)}_{t+1}$ 
and $n_t$ of $\cT^{(3)}_t$.
A mathematically correct statement is 
\beq\label{j11}
 P^{(T)} (n_1, \dots, n_T) \  {\propto} \
 \sum_{\cT_i^{(3)}} 
  \langle \cT_1^{(3)} | \cM | \cT_2^{(3)} \rangle 
  \langle \cT_2^{(3)} | \cM | \cT_3^{(3)} \rangle \dots  \langle \cT_T^{(3)} | \cM | \cT_1^{(3)} \rangle,
 \eeq 
 where the summation  is over all $\cT_i^{(3)}$, $i=1,\ldots T$ satisfying 
 the constraint\footnote{$N_3(\cT^{(3)}_t)$ denotes the number of tetrahedra in the 
 three-dimensional triangulation $ \cT^{(3)}_t$. Recall that by convention we have multiplied
 this number by 2 to obtain $n_t$.}
 $2N_3(\cT_i^{(3)}) = n_i$. The matrix $\la n_{t+1} | M | n_t\ra $ 
 can be thought of as an average 
 over all matrix elements $\la \cT^{(3)}_{t+1} | \cM | \cT^{(3)}_t\ra$ with the constraints
 {$2N_3(\cT^{(3)}_t) = n_t$ and $2N_3(\cT^{(3)}_{t+1}) = n_{t+1}$}. We call $M$ the 
 {\it effective transfer matrix} to distinguish it from the real transfer matrix $\cM$, 
 and the  relation \rf{eq:efftm} is only an approximation and 
 one has to check to what extent it is valid. This was 
 done in detail in \cite{transfermatrix}  and the result was that 
 when the spatial topology was $S^3$ the eq. \rf{eq:efftm} was very well satisfied. We will here assume
 it is also the case when the spatial topology is $T^3$, {the validity of this assumption is discussed} in Section\ \ref{sec:checks}.

The results to be presented below  were obtained 
for four-dimensional CDT with spacetime topology $T^3\times S^1$ (length of $S^1$ being 
two time steps) and the following parameters: $\kappa_0 = 2.2,\ \Delta = 0.6,\ \kappa_4 = 0.9230$.
For technical convenience, a total volume fixing term $\varepsilon(n + m - \bar{c})^2,\ \varepsilon = 10^{-6}$ was added to the bare action, in contrast to the simulations
reported above which were conducted for much longer {length of the periodic time axis} $S^1$. The reason for adding 
this term is to ensure we can collect sufficient statistics for matrix elements 
$\la n | M |m \ra$, where $m,n$ are located in a certain region.
The effect of this volume fixing term is explicitly canceled by hand 
when computing the matrix elements $\la n | M |m \ra$ using  \rf{j10}.
The complete matrix $M$ is then constructed by gluing together the data
obtained for different choices of $\bar{c}$. Here we will merge results from
$16$ patches (simulations) which differ in the value of $\bar{c}$ 
in a range from 100 to 10000. The measured effective  transfer matrix is 
presented in Fig.\ \ref{fig:tmatrix}.

Further, we determine the so-called  effective Lagrangian $L(n, m)$
associated to the effective transfer matrix, defined by 
\beq\label{j12}
\langle n | M | m\rangle = e^{- L(n, m)},\quad S[\{n_t\}] = \sum_t L(n_t, n_{t+1}). 
\eeq
To the extent that \rf{eq:efftm} is valid, the $S[\{ n_t\}]$ defined in \rf{j12} will 
produce the correct probabilities and thus act as an effective 
 action which can be compared to the effective action (\ref{eq:sku}) 
obtained via the covariance matrix method {discussed in Section \ref{sec:cov}.}

There is a significant difference between the effective transfer matrices obtained for
toroidal and spherical topologies. In the spherical case we observed strong discretization
effects when the spatial volume was small. We do not observe similar effects
in the toroidal case. The concave shape shown in Fig.\ \ref{fig:tmatrix} explains
why a volume profile can be localized around some average value,  in contrast
to the spherical case where one needed to fix the spacetime volume during 
the simulations in order to create a non-trivial volume profile. We will discuss the 
reason for this difference below.

\subsection{The kinetic term}

\begin{figure}
\begin{center}
\includegraphics[width=0.96\textwidth]{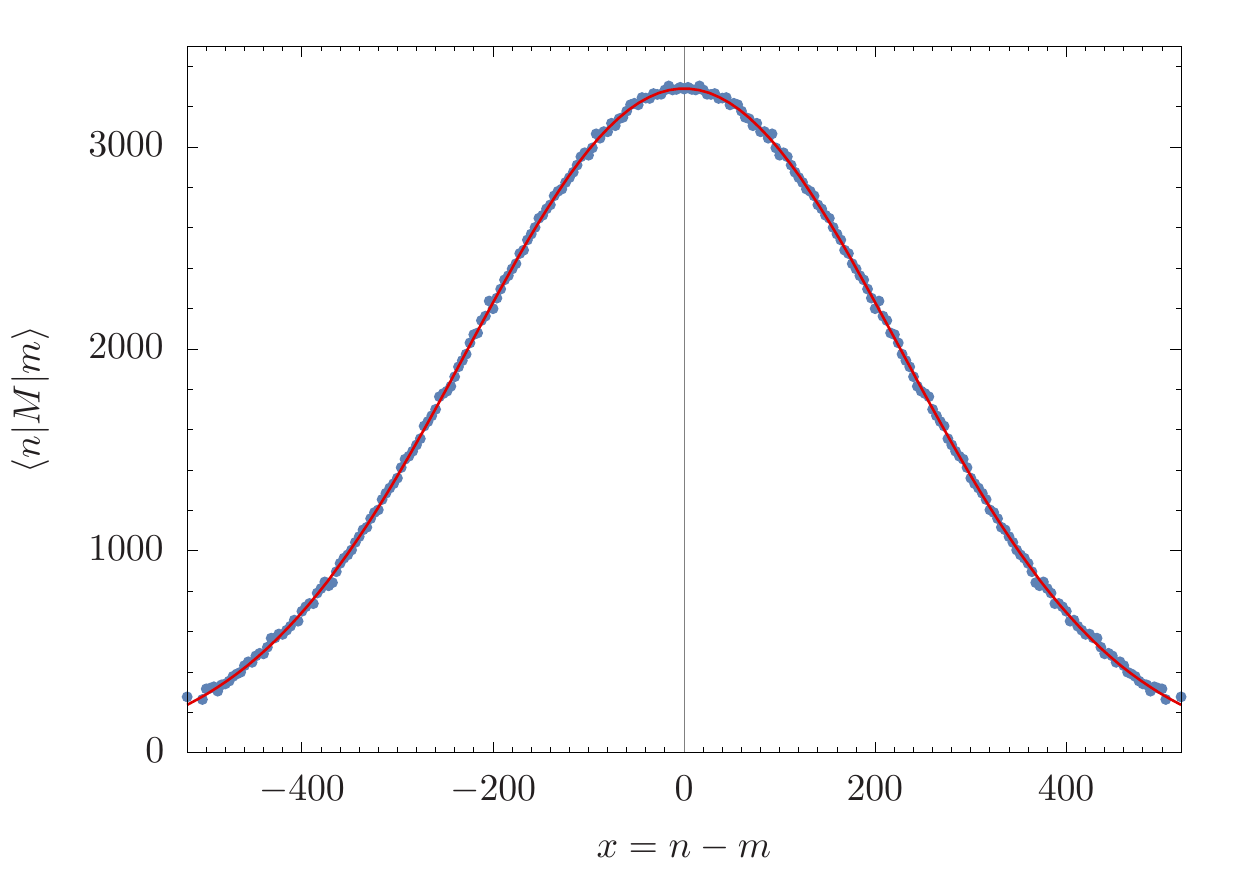}
\end{center}
\caption{Cross-diagonal of the transfer matrix for $c = n + m = 4000$. 
	It matches a Gaussian function  $\cN(c) \, e^{-{x^2/}{f(c)}}$.}
\label{fig:kinanti}
\end{figure}

\begin{figure}
\begin{center}
\includegraphics[width=0.96\textwidth]{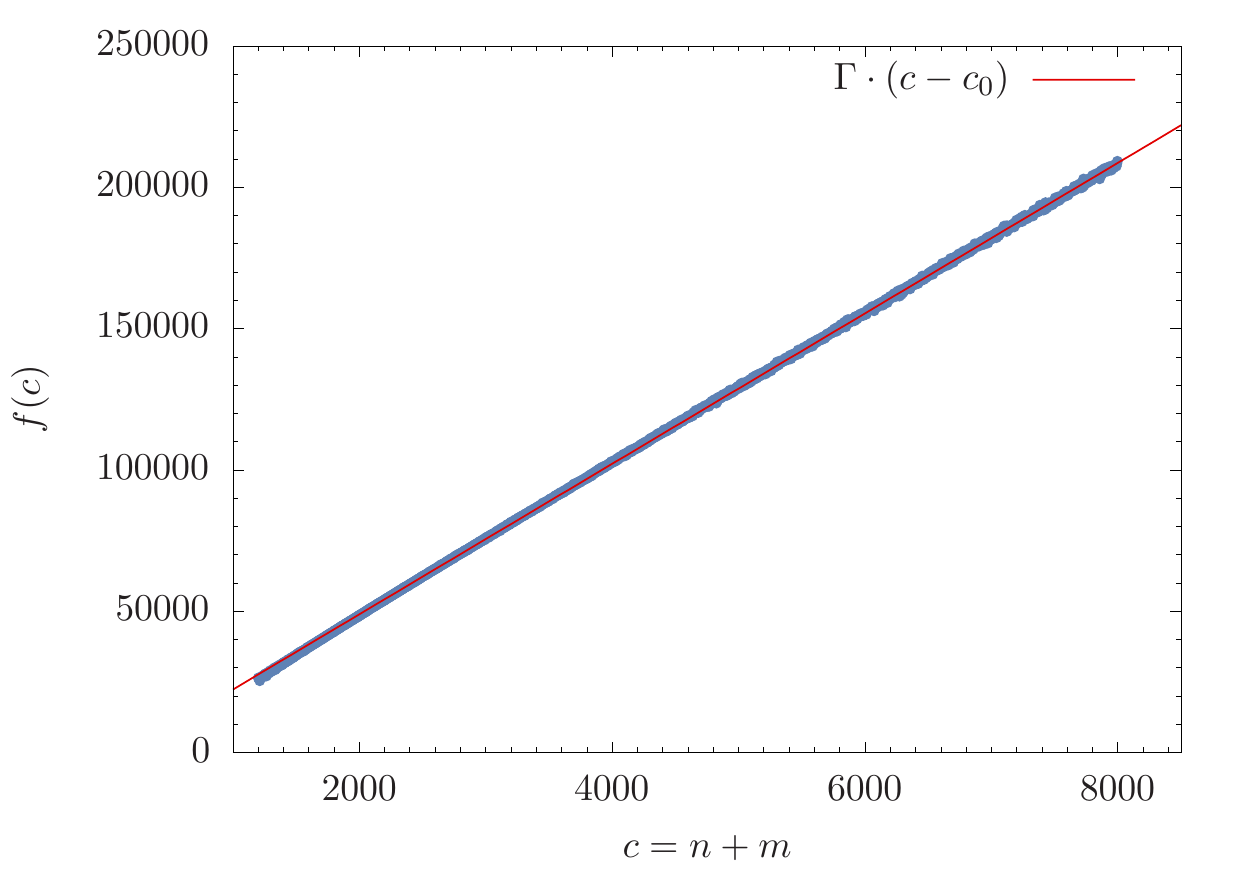}
\end{center}
\caption{Plot of the {kinetic term $f(c)$ extracted by fitting a Gaussian function  $\cN(c) \, e^{-{x^2/}{f(c)}}$ to  
	$\langle n = \frac{c + x}{2} | M | m = \frac{c - x}{2}\rangle$ matrix elements for various values of $c=n+m$ (blue dots) and a linear fit (red line).}}
\label{fig:kinlin}
\end{figure}

We can determine the kinetic part of the effective Lagrangian $L(n,,m)$, i.e.\ the 
part of $L(n,m)$ which depends on the difference $n-m$, by 
analyzing cross-diagonals of the transfer matrix $\langle n | M | m\rangle$ 
{for which $ c= n + m $ is constant}.
An example of such a cross section for $n + m = 4000$ is shown in Fig.\  \ref{fig:kinanti}.
The plot confirms that for constant $c = n + m$ the transfer matrix as a 
function of $x = n - m$ is given with high accuracy by a Gaussian distribution.
Such a distribution  suggests that the transfer matrix elements are described by the formula 
\begin{equation}\label{eq:tmform}
\langle n | M | m \rangle = e^{-\frac{(n - m)^2}{f(n + m)} - U(n + m)}, 
\end{equation}
where $f(c)$ is some function which depends only on the sum $c =n + m$.
The kinetic term $f(c)$ is determined by fitting a Gaussian function 
$\cN(c)\, e^{-x^2 / f(c)}$ to $\langle n = \frac{c + x}{2} | M | m = \frac{c - x}{2}\rangle$ {matrix elements}
for various values of $c$.
A plot of the function $f(c)$ is shown in 
Fig.\ \ref{fig:kinlin}. It is perfectly {well} approximated by a linear fit,
\[ f(c) = \Gamma (c - c_0), \quad \Gamma = 26.61, \quad c_0 = 159.1, \]
which is in agreement with the results obtained in Section \ref{sec:cov:kin},
except for the value $c_0$, which is in both cases small compared to $n+m$, 
and thus not so well determined.

\subsection{The potential term}

The potential $U(c)$ can be extracted from the diagonal {elements} of the transfer matrix.
For $n = m$  the kinetic term gives no contribution and the transfer matrix elements depend only on the potential,
\[ U(c) = - \log \langle n | M | n\rangle = L(n, n), \quad c = 2 n .\]
Instead of directly using  the diagonal elements of matrix $M$ to determine $U(c)$,
another method can be used which significantly reduces statistical errors.
In the previous subsection it was noted that a Gaussian function (\ref{eq:tmform}) fits perfectly to the cross-diagonals of $M$.
Since the fit depends on all points of the cross-diagonal,
the normalization factor $\cN(c) = e^{-U(c)}$ gives a much more accurate estimation of the potential {term}.

\begin{figure}
\begin{center}
	\includegraphics[width=0.96\textwidth]{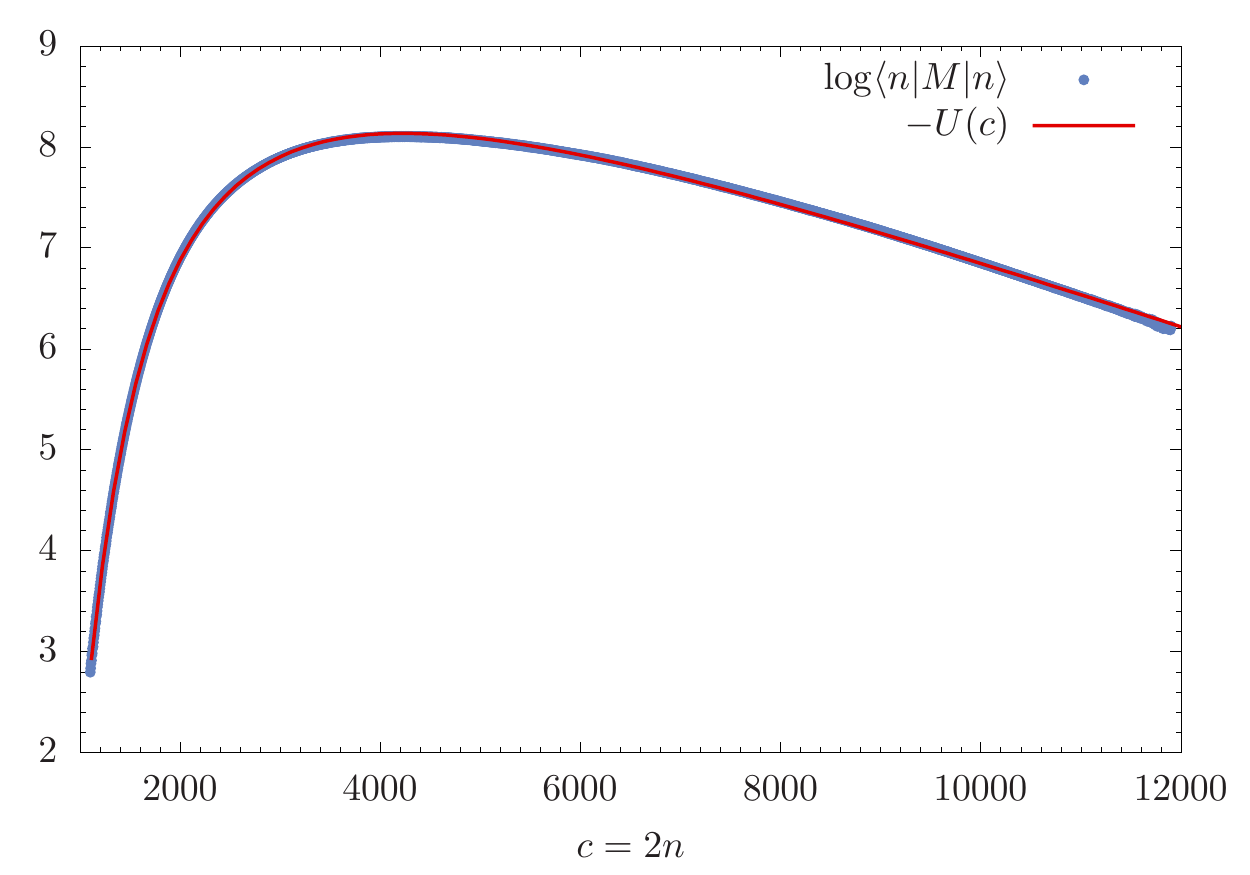}
\end{center}
\caption{
Logarithm of the transfer matrix diagonal $\langle n | M | n \rangle$ 
measured via the normalization factor of the Gaussian fit of the cross-diagonals (blue dots)
and power-law (red line) fit of the potential $U(c)$.}
\label{fig:potential}
\end{figure}

The extracted potential $U(c)$ together with a power-law model is shown in Fig.\ \ref{fig:potential}.
The red line {shows} the best fit power-law function,
\beql{fitU}
U(c) = A + \lambda c + \mu c^\gamma= A + 0.00035\, c + \left( \frac{c}{4132} \right)^{-1.509}.
\eeq
The model fits well the empirical data, with the exponent very close to 
$\gamma = - \frac{3}{2}$,
which is in agreement with the results obtained using  the covariance matrix,
as reported in Section \ref{sec:cov:pot}.
We tried fitting several functional forms {including} a logarithmic function, 
but none of them was as accurate {as the fit \rf{fitU}}.
The value of the exponent $\gamma$ depends on the fit range, 
and varies from $-1.1$ to $-1.6$ which suggest that a power-law behavior is merely a leading order approximation of some more general function.

\subsection{The effective action}
	
Based on the previous results we can state that the effective discrete Lagrangian 
is well approximated  by the following formula:
\begin{equation}
L(n, m) = \frac{(n - m)^2}{\Gamma (n + m - c_0)} + \mu (n + m)^{\gamma} + \lambda (n + m) . 
\label{eq:mod1}
\end{equation}
However, by studying the diagonal itself we cannot distinguish
between the following two versions of the potential, $\mu (n + m)^\gamma$ and $\mu (n^\gamma + m^\gamma)$.
In the latter case, which was used in section \ref{sec:cov}, 
the transfer matrix elements are given by
\begin{equation}
L(n, m) = \frac{(n - m)^2}{\Gamma (n + m - c_0)} + \mu \left(n^{\gamma} + m^{\gamma}\right) + \lambda (n + m) . 
\label{eq:mod2}
\end{equation}
To decide which form of the potential is more suitable,
we compare the models (\ref{eq:mod1}) and (\ref{eq:mod2}) 
with the total effective  transfer matrix
and calculate the fit goodness {
\[ E = \sum_{n,m} \left( \langle n | M | m\rangle - e^{- L(n, m)} \right)^2 / \sum_{n,m} \langle n | M | m\rangle^2,\]} 
where the sum is over all measured elements of the transfer matrix and $L(n, m)$ is the considered model.
The two models give very comparable results.
The best fit of model (\ref{eq:mod1}) gives
$\Gamma = 26.86, \ c_0 = 182.4, \ \gamma = -1.49945, \ \mu=4178^{-\gamma}, \ \lambda = 3.47 \cdot 10^{-4}, \ E = 0.55 \cdot 10^{-3}$,
the second model (\ref{eq:mod2}) gives 
$\Gamma = 26.22, \ c_0 = - 6.75, \ \gamma = -1.44757, \ \mu=1328^{-\gamma}, \ \lambda = 3.47 \cdot 10^{-4}, \ E = 0.46 \cdot 10^{-3}$.
The latter version has slightly smaller deviation $E$ but nothing which can serve as 
a motivation to prefer  \rf{eq:mod2} to \rf{eq:mod1}.

To summarize: measurement of the effective transfer matrix for {the toroidal spatial topology}
within phase {$C_{dS}$} of four-dimensional  CDT ($\kappa_0 = 2.2, \Delta = 0.6, \kappa_4 = 0.9230, T = 2$)
shows
that the transfer matrix elements are quite precisely expressed by the following formula
\begin{equation}
\langle n | M | m\rangle = e^{- \left[ \frac{(n - m)^2}{\Gamma (n + m - c_0)} + \mu (n + m)^{-3/2} + \lambda (n + m)\right]} . 
\end{equation}

\section{{Checks of the effective matrix model}}\label{sec:checks}

\begin{figure}
\begin{center}
\includegraphics[width=0.9\textwidth]{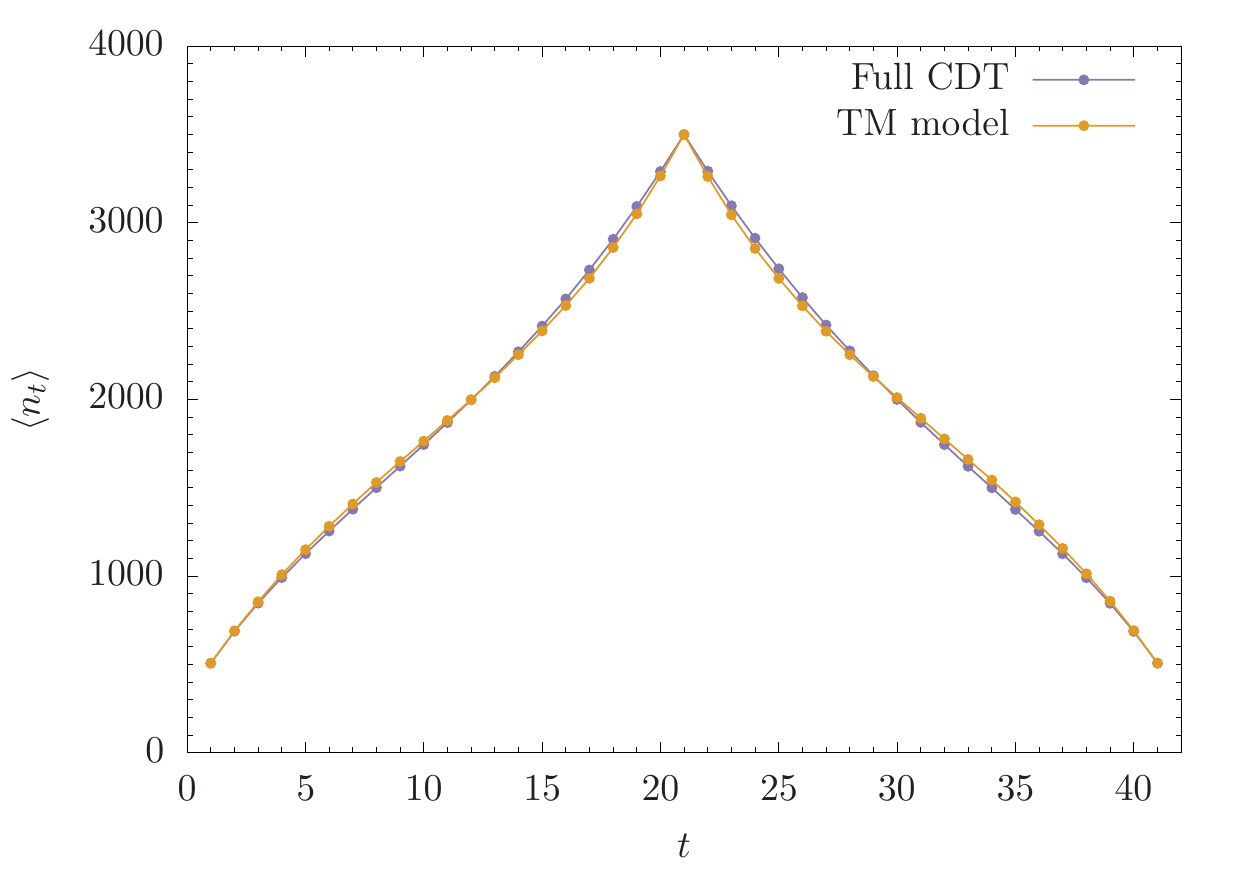}
\end{center}
\caption{Average volume profile $\langle n_t \rangle$ obtained for the full CDT model and using the effective transfer matrix model.}
\label{fig:avntmcv}
\end{figure}

In Sections \ref{sec:cov} and \ref{sec:classical} we studied {the effective action extracted from}
the covariance matrix for spatial volumes 
obtained from CDT simulations with toroidal topology and $T = 40$ time slices.
In Section \ref{sec:tm} the empirical transfer matrix, measured for $T = 2$ time slices, was used to extract the effective action.
Here, we check if the decomposition (\ref{eq:efftm})
\[ P^{(T)} (n_1, \dots, n_T) \ {\propto} \ \langle n_1 | M | n_2 \rangle \langle n_2 | M | n_3 \rangle \dots  \langle n_T | M | n_1 \rangle \]
can reproduce the full CDT results for an arbitrary number of slices $T$,
and thus verify whether the effective transfer matrix approach is legitimate.
{We reduce the problem to the, so called, balls-in-boxes model \cite{bib},
where the only dynamical degrees of freedom are spatial volumes $n_t$.}
The procedure is as follows: we {use} 
the effective transfer matrix $M$ measured in Section \ref{sec:tm} for $T = 2$ 
{to generate spatial volume profiles with $T = 40$ time slices.}
To make the setup identical {to CDT simulations}, the same volume fixing terms as in (\ref{eq:sfix}) are included.
Using the Metropolis algorithm a {large set of} volume {distributions} $\{n_t\}$ is generated
according to probability (\ref{eq:efftm}).
Further, we calculate the average volume  profile $\langle n_t \rangle$ and {the amplitude of volume fluctuations  $\Delta n_t =\sqrt{ \langle (n_t - \bar{n}_t)^2 \rangle}$},
and compare it with the results of the full CDT model with $T = 40$.

\begin{figure}
\begin{center}
\includegraphics[width=0.9\textwidth]{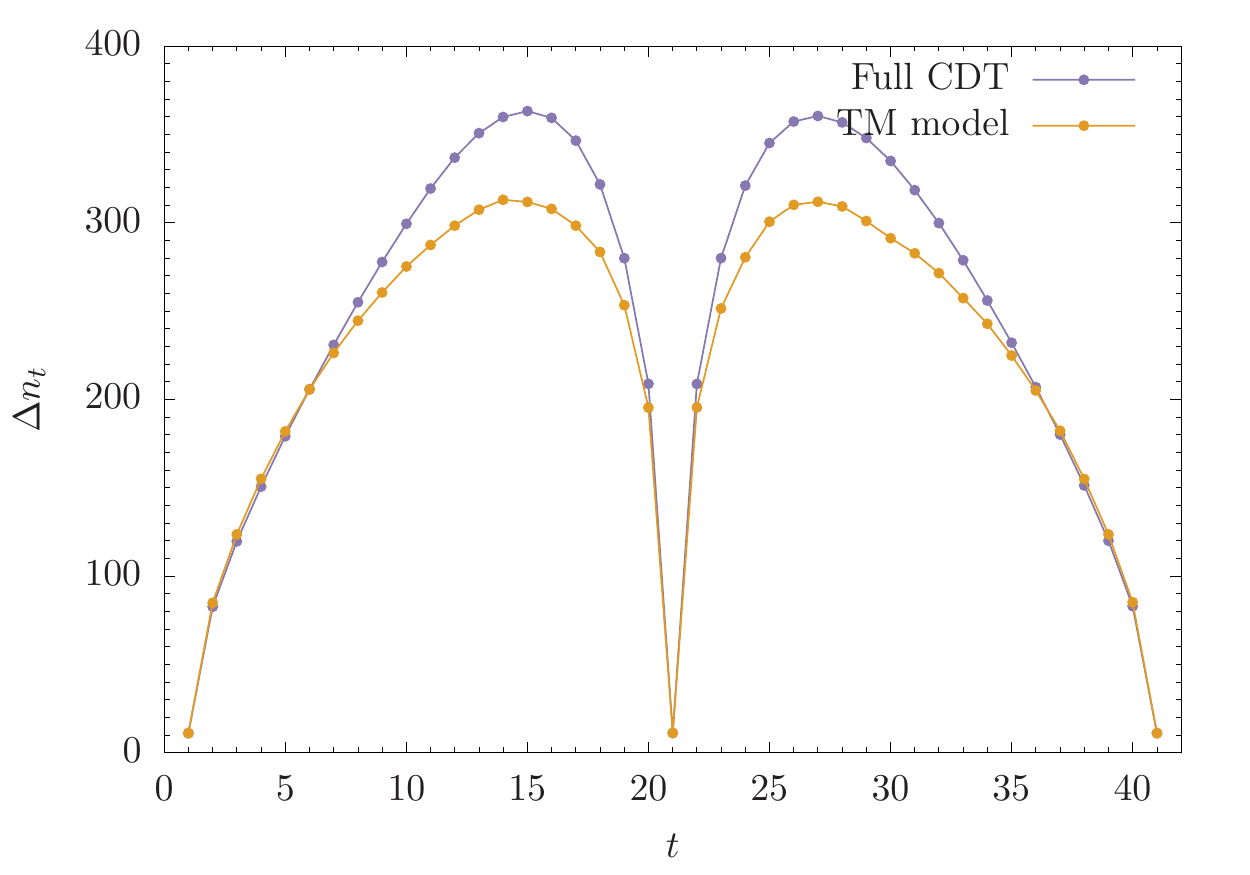}
\end{center}
\caption{Amplitude of fluctuations of the spatial volume $\Delta n_t$ obtained for the full CDT model and using the effective transfer matrix model.}
\label{fig:varntmcv}
\end{figure}

A plot of the average volume profile $\langle n_t \rangle$ is shown in Fig. \ref{fig:avntmcv}.
The yellow curve was obtained from the balls-in-boxes  model (\ref{eq:efftm}) based on the the effective transfer matrix $M$ measured in four-dimensional CDT simulations with spacetime topology $T^3\times S^1$
and $T = 2$ time steps and the following parameters: $\kappa_0 = 2.2,\ \Delta = 0.6,\ \kappa_4 = 0.9230$.
The blue curve, which is the same as in Fig. \ref{fig:average},
plots $\langle n_t \rangle$ measured 
in full CDT simulations {with $T = 40$ time steps}  for the same values of bare coupling constants $\kappa_0$ and $\Delta$,
but    {slightly different}  $\kappa_4 = 0.9225$.
The two curves overlap almost exactly.
However, {one should note that if in the balls-in-boxes simulations one used the effective transfer matrix $M$ measured for $\kappa_4 = 0.9225$, i.e. exactly the same as in the full-CDT simulations, one would obtain a
"bulgy" volume profile $\langle n_t \rangle$ which would not reproduce the full CDT results.}
The amplitude of spatial volume fluctuations $\Delta n_t$ 
for the {balls-in-boxes} 
model (yellow curve, $\kappa_4 = 0.9230$) and 
the full CDT model (blue curve, $\kappa_4 = 0.9225$)
is shown in Fig. \ref{fig:varntmcv}.
Although the two curves are similar, the discrepancy cannot be removed simply by tuning only $\kappa_4$.

The results for the effective transfer matrix approach
and the full CDT model are qualitatively equivalent,
but the small discrepancy suggests that the effective transfer matrix $M$ depends on the number of spatial slices $T$.
{Probably}, it can be levelled by introducing a dependence of bare coupling constants on $T$.
We have shown that a slight change of $\kappa_4$ recovers the average volume profile. 
Failure to exactly reproduce the full CDT results using an effective description via spatial volumes $n_t$
implies that some perturbations propagate through the geometrical structure of spatial slices
and introduce long-range correlations which cannot be neglected.
The reason why a disagreement between the effective transfer matrix model
and the full CDT model is more pronounced in the toroidal case than in the spherical case {\cite{transfermatrix}},
might be due to the lack of a {semiclassical} potential term
in the effective action. {Such a potential $\propto n_t^{1/3}$ is predicted by the minisuperspace model in the spherical  case due to  non-vanishing curvature of the spherical minisuperspace solution, and it is not present in the toroidal case where the minisuperspace reduction yields a flat spacetime}.
In the spherical case, the minisuperspace reduction outlines a leading behaviour of spatial {volumes}
and their internal structure is less {visible} in the effective description.

\section{Discussion}

In CDT one chooses by hand the  topology of space and insists on a time-foliation.
One Wick rotates spacetime to Euclidean signature in order to make Monte Carlo simulations
possible, and imposes periodic boundary conditions in Wick-rotated time for convenience in
the simulations. The {bare} action used is  the standard Regge 
action. Extensive studies have been performed in the case where the topology of space 
was $\Sigma=S^3$ and a clear picture emerges if one only looks at spatial volume as a function 
of the foliation time: there is a background spatial volume profile $\bar n_t$ around which 
there are well defined quantum fluctuations. The background profile, as well as the 
quantum fluctuations are well described by an effective minisuperspace action.
In contrast to the standard minisuperspace action, this one is not obtained by {imposing strict symmetries on the metric field, i.e. by demanding that }  possible fluctuations of the geometries  depend only
on the scale factor $a(t) \propto n_t^{1/3}$, but is {rather} obtained by integrating out  
other degrees of freedom { in a full lattice model with no geometric symmetries except from spatial topology imposed}. Although it superficially appears similar to the standard 
minisuperspace action, it is in fact radically different, since the ``kinetic" term has  
{opposite sign compared to the standard minisuperspace reduction}. Thus the entropy of configurations, i.e.\ the measure of the path integral,
plays an important role {as it dynamically corrects the wrong sign of the minisuperspace action}.$^1$ For the {spherical  topology of spatial slices the form of the effective action}  
 was as follows
\begin{equation}\label{j20}
{S^{\Sigma=S^3}[\{n_t\}] = \sum_t\left[ \frac{1}{\Gamma} \frac{(n_{t + 1} - n_t)^2}{n_{t + 1} + n_t} + 
\alpha \,n_t^{1/3}+ \mu \, n_t^{\gamma} - \lambda \, n_t\right] .}
\end{equation}
 The term $n^{1/3}_t$ was interpreted as being the equivalent of the classical 
 term present in the ordinary minisuperspace action  while the term $n_t^{\gamma}$ 
 was generated by quantum corrections. For the spherical case this term could not be 
 reliably determined {due to its sub-leading character}. 
 
 If the spatial topology is that of $\Sigma=T^3$ the situation is somewhat different. The geometry
 fluctuates around a  different  {semiclassical background}. One can again try to determine the
 {effective} action which describes the volume profile and its fluctuations {and to compare it with the minisuperspace reduction for a regular torus}. In this 
 article we have done this in two different ways, 1/ by using the measured 
  volume-volume {covariance matrix}  and 2/ by measuring the effective 
 transfer matrix. The results agree within the statistical errors of the measurements.
 One finds {that}
  \begin{equation}\label{j21}{
S^{\Sigma=T^3}[\{n_t\}] = \sum_t \left[\frac{1}{\Gamma} \frac{(n_{t + 1} - n_t)^2}{n_{t + 1} + n_t} + 
\mu \, n_t^{\gamma} + \lambda \, n_t \right]}.
\end{equation} 
 The dominant term (from a numerical point of view) in both 
 \rf{j20} and \rf{j21} is the kinetic term. Both terms have the same form 
 and also the coefficient $\Gamma$ agrees. We {conjecture} that this term most likely is  
 universal {in CDT}, independent of the {spatial topology chosen}.\footnote{{Similar kinetic term was also observed for a two- and three-dimensional CDT.}} In the toroidal case 
 there is no ``classical'' $n_t^{1/3}$ term. This is in agreement with the minisuperspace
 {reduction as such a term observed in spherical case was classically due to  positive curvature of the sphere and it should not be present for a (flat) geometry of a torus}. In the case of the sphere it was 
 difficult to determine in a reliable way the quantum correction term 
 $n_t^{\gamma}$ because it was subdominant compared to  the term $n_t^{1/3}$. 
 However, for the torus, where this term is absent, we can actually determine 
 a correction term. We find that $\gamma$ is close to the value $-3/2$, but there are 
 indications that such a power law is only an approximation to {a more complicated}
 function. Presently we have no simple explanation {for this quantum correction} even if the power
 law should turn out to be exactly $-3/2$. Also, we do not know if this term is  
 universal  independent of the spatial topology chosen since, as mentioned, we have 
 not been able to determine it in the spherical case. 
 
In summary, CDT provides a theory of fluctuating four-dimensional
geometries which, if we only look at the scale factor, allows a description 
in terms of a minisuperspace action which has the kinetic and potential 
terms one would expect for the given spatial topology. {It is quite intriguing that a simple minisuperspace reduction seems to  explain spatial volume data of a system with no geometric symmetries put in by hand so well even for very small "universes" dominated by quantum fluctuations.\footnote{{Estimated  radius of a typical CDT "universe" is of the order of a few Planck lengths.}} This observation can shade some light in favour for a validity of spatially isotropic and homogenous models commonly used in cosmology.}  In the case of the torus
studied here, since the classical potential term is absent, we could also 
observe a quantum correction term. It would be interesting if it could be 
also obtained by analytic calculations.

\section*{Acknowledgements}

A.G. acknowledges support by the National Science Centre, Poland under 
Grant No. 2015/17/D/ST2/03479. J.A. and K.G. acknowledge the support by  the 
ERC-Advance grant 291092, ``Exploring the Quantum Universe'' (EQU).
{The research of J.A. was supported in part by Perimeter Institute of Theoretical Physics. Research
of the Perimeter Institute is supported by the Government of Canada and by the Province of Ontario.}
{ J.G-S. and J.J. wish to acknowledge the support of the grant DEC-2012/06/A/ST2/00389 from the National Science Centre, Poland. }


%



\end{document}